\documentclass{gGAF2e} 

\usepackage{graphicx}
\usepackage{dcolumn}
\usepackage{bm}
\usepackage{lineno}
\usepackage{color}
\usepackage{hyperref}

\begin{document}

\title{Exciting Standing Rossby Waves in the Large Rotating Annulus}

\author{
Kial D. Stewart${^{1,2,3}}$$^{\ast}$\thanks{$^\ast$Corresponding author. Email: kial.stewart@anu.edu.au}, Callum J. Shakespeare${^{1,2}}$ \& Thomas G. Schmaltz${^{2,4,5}}$\\
${^1}$ Climate \& Fluid Physics Laboratory, Australian National University\\
${^2}$ ARC Centre of Excellence for Climate Extremes, University of New South Wales, Australia\\
${^3}$ ARC Centre of Excellence for 21st Century Weather, Monash University, Australia\\
${^4}$ Climate Change Research Centre, University of New South Wales, Australia\\
${^5}$ ARC Centre for Excellence in Antarctic Science, University of Tasmania, Australia}

\maketitle

\linespread{1.5}
\begin{abstract}
\section*{Abstract}

Laboratory experiments with rotating tanks remain the premier physical analogue for atmospheric dynamics.
Often, the equipment involved is engineered to be sufficiently versatile and modular so as to be able to accommodate experiments that explore a wide range of atmospheric processes.
The exercise of initially configuring the apparatus then involves running experiments that sweep through parameter space to identify the specific dynamical regimes of interest.
This process is typically considered part of the development and testing phase of a given project, and these initial experiments are generally left unreported; while many of these test experiments may not be directly relevant to the project at hand, they may be useful for other applications or scientific communities.
Here we report on a series of 93 different laboratory experiments run with the intention of identifying suitable experimental configurations that excite and sustain standing Rossby waves.
We use the Large Rotating Annulus at the Australian National University; this unique apparatus has independent control over the annulus sidewall temperatures and background rotation rate, as well as the rotation rate of a differentially-rotated topographic bump.
These three conditions are systematically varied through an extensive range of experimental parameter space.
Configurations are found in which standing Rossby waves are present; unsurprisingly, these are experiments with eastward zonal flow and a sufficiently strong background gradient of potential vorticity.
The behaviour of these standing Rossby waves follows that predicted by established theory; for example, the wavelength of the standing waves scales with the square root of the background flow speed divided by the gradient of potential vorticity.
The standing and transient variability fields are quantified, and the partitioning of variability between them depends on the existence of standing Rossby waves; the variability in experiments with and without standing Rossby waves is predominantly standing and transient, respectively.
Subsequent efforts will focus on specific experiments with standing Rossby waves, with the aim to provide insight into how wave behaviour may respond to climate change.

\end{abstract}

\begin{keywords}

differentially-heated rotating annulus; Rossby waves

\end{keywords}

\section{Introduction}

Laboratory experiments with rotating tanks have long been a powerful and effective means to investigate many aspects of geophysical and astrophysical fluid dynamics \cite[e.g.,][]{taylor1921, hide1958, read2019, harlander_etal2024}.
Their simple geometries and boundary conditions can be well described analytically, and provide rigorous testbeds for theories relating to, but not limited to, barotropic and baroclinic instabilities, geostrophic and zonostrophic turbulence, rotating convection, Rossby waves and jets; that is, fluid processes that are challenging, expensive, or infeasible to examine with alternative approaches.
Recent developments in diagnostic technologies and analysis techniques have renewed interest in rotating fluid experiments, granting new insights and understanding into fundamental flow dynamics, and guiding the development of parameterisations to better represent these processes in modern numerical simulations.
Experiencing rotating fluid dynamics firsthand in laboratory settings has enormous pedagogical utility and benefits, especially for building intuition and novel perspectives; these experiences are impossible to achieve with digital representations of fluid systems.
Indeed, the lamentation of \cite{taylor1921} rings eternal, and with new context in the supercomputer age: ``It is well known that predictions about fluid motion based on the classic hydrodynamical theory are seldom verified in experiments performed with actual fluids.''

The intended scientific application generally determines the specific configuration of the rotating apparatus and boundary conditions; for example, experiments designed to study the atmospheric dynamics of gas giants \cite[e.g.,][]{condie_rhines1994, aguiar_etal2010, cabanes_etal2017} are configured differently to those that explore barotropic zonal jets \cite[e.g.,][]{sommeria_etal1989, weeks_etal1997, tian_etal2001, stewart_macleod2022}.
The ``Hide Tank'' configuration, named after \cite{hide1958}, is a versatile approach which has recently been used to explore a range of dynamics including inertia-gravity waves \cite[e.g.,][]{rodda_etal2018, rodda_etal2020, rodda_harlander2020}, oceanic jets \cite[e.g.,][]{smith_etal2014}, and rotating stratified flow-topography interactions \cite[e.g.,][]{stewart_shakespeare2024}.
A Hide Tank consists of a rotating annulus in which the inner and outer sidewalls are maintained at different temperatures, thereby providing an environment in which a stratification can be maintained indefinitely.
The differentially-heated sidewalls drive a large-scale convective overturning circulation that is affected by the background rotation of the annulus, leading to a predominantly zonal flow that can exhibit dynamics equivalent to Rossby waves, gyres, jets, and baroclinic geostrophic turbulence \cite[e.g.,][]{wordsworth_etal2008, rodda_etal2020, rodda_etal2022, harlander_etal2022}.
The background gradient in potential vorticity, which on Earth is primarily due to the latitudinal gradient of the Coriolis parameter, and is a necessary condition for supporting Rossby waves, can be represented in rotating tank experiments by way of a spatially-dependent fluid depth; this can either be engineered into the apparatus as a sloping lid and/or base \cite[e.g.,][]{wordsworth_etal2008, smith_etal2014, stewart_macleod2022}, or achieved dynamically by rotating the annulus at a sufficiently rapid rate \cite[e.g.,][]{cabanes_etal2017}.
This means that a Hide Tank configuration which combines a large-scale convective overturning circulation, stratified geostrophic turbulence, and a background gradient in potential vorticity, conceivably has the necessary features to serve as an idealised domain for modelling mid-latitude atmospheric dynamics, such as Rossby waves.
Thus, if we take mid-latitude atmospheric Rossby waves to be the intended scientific application for our Hide Tank experiments, the challenge then is to ensure that the experiments are performed in a suitable dynamical regime that maximises their relevance to this application, which typically involves extensive testing and tuning of the apparatus.

A traditional Hide Tank has several degrees of freedom: the inner and outer radii of the annulus, depth of the working fluid, the rotation rate of the annulus, the inner and outer sidewall temperatures, and the properties of the working fluid itself.
Some of these parameters are relatively simple to change, like the rotation rate of the annulus, while others are more challenging, like the geometry of the annulus.
Typically, a series of experiments with Hide Tanks will aim to hold many of these parameters constant such that the system is in a dynamically-relevant regime, and vary one parameter while measuring the response.
That said, the initial exercise of identifying the optimal experimental configuration for the dynamically-relevant regime of interest is often considered part of the apparatus development and testing process, and thus ignored in the scientific literature.
This unfortunate practice leaves a plethora of science unreported; while these particular trial experiments may not be directly applicable to the initial intended purpose of the project at hand, they may be useful for other applications or scientific communities.

Here we present a comprehensive series of laboratory experiments in a Hide Tank using the Large Rotating Annulus (LRA) in the Climate \& Fluid Physics Laboratory at the Australian National University.
These experiments explore a wide range of parameter space with the intention of identifying a dynamical regime that is optimal for mid-latitude atmospheric dynamics; specifically, standing Rossby waves.
We employ the differentially-rotating topography system described by \cite{stewart_shakespeare2024} to rotate a topographic bump around the LRA and generate flow-topography interactions.
We vary three experimental parameters; the rotation rate of the LRA, the sidewall temperature difference, and the differential rotation rate of the bump.
Standing Rossby waves are excited in cases that have a sufficiently rapid LRA rotation rate and a bump rotation that corresponds to eastward flow over the topography.
The magnitude and sign of the sidewall temperature difference affect the baroclinicity of the experiments.
The layout of the paper is as follows: in \S2 we cover some theoretical background relevant to Rossby waves in Hide Tanks; in \S3 we describe the apparatus, methodology, and analysis techniques; in \S4 we present and discuss the experimental results, and provide our conclusions in \S5.

\section{Theoretical Background}

In this section we introduce some theoretical considerations that are important for Rossby waves in rotating annular systems like Hide Tanks.
We first introduce the geometry of the apparatus, the processes by which it can support a background gradient in the potential vorticity and, subsequently, Rossby waves.
Then we consider the effects that a sidewall temperature difference can have in generating stratified flow dynamics, and how these relate to previous experiments with Hide Tanks.
Finally, we comment on the physical implications that the sign of the sidewall temperature difference can have in terms of the baroclinicity of the system.

\begin{figure}[]
\centering
\includegraphics[width=0.7\textwidth]{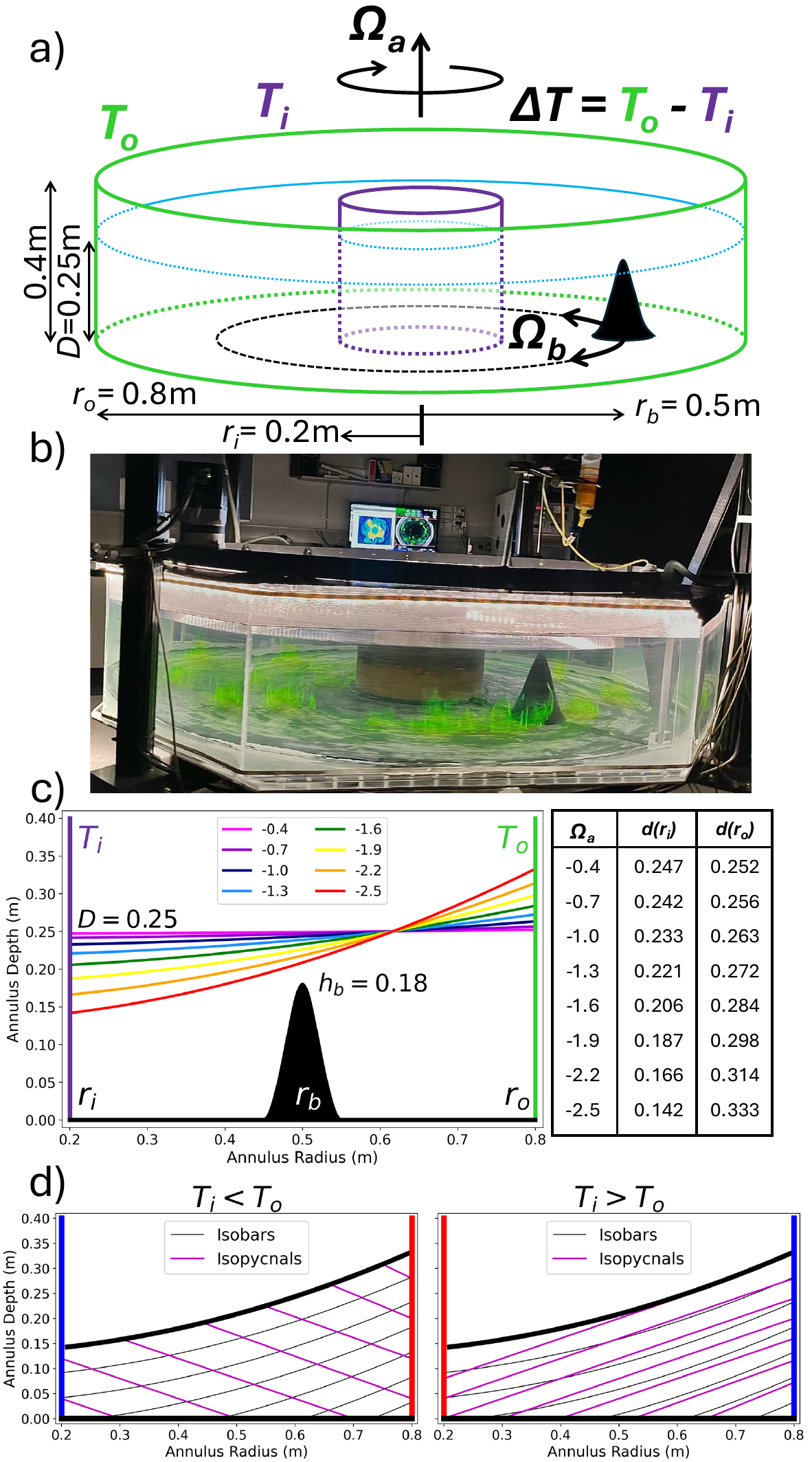}
\caption{Aspects of the experimental setup. Panel (a) shows a schematic of the Large Rotating Annulus (LRA) and differentially-rotating topographic bump. Panel (b) is a photo of the LRA during a qualitative experiment; in practice, the outer sidewalls are insulated and heavy curtains completely enshroud the LRA. Panel (c) is a schematic of a radial transect across the annulus showing the free surface of the working fluid for different annulus rotation rates $\Omega_{a}$ (rad/s; Eqn. \ref{eqn:d_by_r_Omega}); the associated table lists the depth (in m) of the fluid at the inner and outer sidewalls. Panel (d) is a schematic depicting the effect that the sign of the sidewall temperature difference $\Delta{T}$ has on the isopycnal slope; positive $\Delta{T}$ has isopycnals that slope against the isobars, while negative $\Delta{T}$ has isopycnals that slope in the same direction as the isobars.}
\label{fig:01_LRA_schematic}
\end{figure}

\subsection{Geophysical Fluid Dynamics in a Rotating Annulus}

Consider an incompressible fluid of reference density $\rho_{0}$ (kg/m$^3$) in an annulus with inner and outer sidewall radii $r_{i}$ and $r_{o}$ (m), respectively, and of radial width $L = r_{o} -r_{i}$ (m) and reference depth $D$ (m; Fig. \ref{fig:01_LRA_schematic}a,b).
The annulus is able to rotate about its central vertical axis at a steady rate of $\Omega_{a}$ (rad/s), where the Coriolis parameter is $f=2\Omega_{a}$.
The dynamical influence of the background rotation on the system can be represented by the ratio of the rotational timescale ($1/2\Omega_{a}$) and advective timescale ($L/U$), as,
\begin{equation}
Ro = \frac{U}{2\Omega_{a}{L}} =  \frac{U}{f{L}} ,
\end{equation}
where $U$ and $L$ are characteristic velocity and length scales (represented by the radial width of the annulus), respectively; the system is sensitive to the background annulus rotation when the Rossby number is much smaller than unity, $Ro<<1$.

The background rotation of the annulus causes the contained fluid to experience a radially-outwards centrifugal force that scales with the product of the rotation rate squared, fluid density, and the local radius, $\Omega_{a}^{2}\rho_{0}{r}$.
This radially-outwards centrifugal force results in the fluid tending to accumulate near to the outer sidewall, thereby increasing the depth of the fluid $d$ at the outer regions of the annulus, and leading to a hydrostatic pressure gradient oriented radially-inwards, $\partial{P}/\partial{r}$, where the hydrostatic pressure is defined as $P=\rho_{0}{g}{d}+C$ for gravity $g$ and integration constant $C$.
In time, the outward centrifugal force and the inward hydrostatic pressure gradient come to balance;
\begin{equation}
\Omega_{a}^{2}\rho_{0}{r} = -\frac{\partial{P}}{\partial{r}},
\end{equation}
which reduces to,
\begin{equation}
\frac{\partial{d}}{\partial{r}} = \frac{\Omega_{a}^{2}r}{g}.
\end{equation}
This expression can be integrated in $r$ between the limits of $r_{i}$ and $r_{o}$ and combined with the reference depth $D$ to return the depth of the fluid as a function of radius $r$ and annulus rotation rate $\Omega_{a}$,
\begin{equation}
d(r,\Omega_{a}) = D -\frac{\Omega_{a}^{2}}{4g}\left(r^{2}_{o} + 3r^{2}_{i} - 2r^{2}\right).
\label{eqn:d_by_r_Omega}
\end{equation}
That is, when the radially-outwards centrifugal force and the radially-inwards hydrostatic pressure gradient are in balance, the fluid depth obtains a parabolic shape that depends on the radius squared (e.g., Fig. \ref{fig:01_LRA_schematic}c).

Potential vorticity $q$ is a conserved property of fluids, and is defined as,
\begin{equation}
q = \frac{f+\zeta}{d},
\end{equation}
where $\zeta=\partial{v}/\partial{x}-\partial{u}/\partial{y}$ is the relative vorticity of the flow.
A scaling for the relative vorticity $\zeta$ can be given by the ratio of the characteristic velocity and length scales, $\zeta\sim{U/L}$; this scaling provides an alternative definition for the Rossby number as,
\begin{equation}
Ro = \frac{\zeta}{f}.
\end{equation}
Thus, for a system that is sensitive to the background rotation (e.g., $Ro<<1$), the small Rossby number implies $\zeta<<f$, such that the potential vorticity $q$ can be approximated by $q\approx{f}/d$.

The latitudinal dependence of the Coriolis parameter $f$ results in a latitudinal gradient of potential vorticity, which is typically linearised about a reference Coriolis value $f_{0}$ and represented as $\beta$
\begin{equation}
f \approx f_{0} + \beta{y} \quad\rightarrow\quad \frac{\partial{f}}{\partial{y}}=\beta.
\end{equation}
This approximation requires an assumption that the fluid system is ``thin'' in the sense that the horizontal lengthscale $L$ is much larger than the vertical lengthscale $D$, $L>>D$, such that the latitudinal gradient in the Coriolis parameter is larger than the latitudinal gradient in fluid depth.

In the case of the rotating annulus, there is a constant and uniform value of $f=2\Omega_{a}$.
There is however a radial dependence of fluid depth (e.g., Eqn. \ref{eqn:d_by_r_Omega}) which can support a background radial gradient of potential vorticity $q$ that is dynamically equivalent to that given by the latitudinal gradient of $f$.
Considering that $f$ and $d$ are directly and inversely related to $q$, respectively, an increase in fluid depth $d$ is dynamically equivalent to a decrease in the magnitude of Coriolis parameter $f$ (i.e., accounting for when $f<0$), and vice versa.
Thus, the radial dependence of fluid depth in the rotating annulus can be expressed as a radially-dependent effective Coriolis parameter,
\begin{equation}
f \approx f_{0} + \beta{r} \quad\rightarrow\quad \frac{\partial{f}}{\partial{r}}=\beta.
\label{eqn:effective_f}
\end{equation}
Here, $\beta$ has the form of the Coriolis parameter scaled by the radial gradient in fluid depth normalised by the reference depth;
\begin{equation}
\beta = \frac{f}{D}\frac{\partial{d}}{\partial{r}} = \frac{2\Omega_{a}^{3}r}{gD}.
\end{equation}
Using the mid-annulus radius $(r_{o}+r_{i})/2$, a characteristic value of $\beta$ as a function of annulus rotation rate $\Omega_{a}$ becomes,
\begin{equation}
\beta(\Omega_{a}) = \frac{\Omega_{a}^{3}\left({r_{o}+r_{i}}\right)}{gD}.
\label{eqn:beta}
\end{equation}

The relative level of influence that $\beta$ will have on the system can be represented by two dimensionless terms,
\begin{equation}
\beta_{1} = \frac{\beta{L}}{f},
\label{eqn:beta1}
\end{equation}
and
\begin{equation}
\beta_{2} = \frac{\beta{L^{2}}}{U},
\label{eqn:beta2}
\end{equation}
referred to as the planetary number and planetary-relative vorticity number, respectively.
For $\beta_{1}<<1$, the effective Coriolis parameter is relatively uniform throughout the system and the influence of $\beta$ is negligible; that is, the system is dominated by $f$-plane dynamics as opposed to $\beta$-plane dynamics.
The second term $\beta_{2}$ provides a comparison of $\beta$ to the relative vorticity gradient (which scales with $U/L^{2}$).

The fact that the rotating annulus can support a background radial gradient of potential vorticity means that Rossby waves can occur and propagate within the system.
Following \cite{pedlosky1987}, these Rossby waves have a dispersion relation given by,
\begin{equation}
\omega = \frac{-\beta{L_{R}^{2}}k}{1+L_{R}^{2}\left(k^{2}+l^{2}\right)}
\label{eqn:rossby_dispersion}
\end{equation}
where $\omega$ is the Rossby wave frequency, $k$ and $l$ are the zonal and meridional wave numbers, respectively, which relate to the zonal and meridional wavelengths ($\lambda_{k}$, $\lambda_{l}$, respectively) by $k=2\pi/\lambda_{k}$ and  $l=2\pi/\lambda_{l}$, respectively, and $L_{R}$ is the barotropic Rossby radius of deformation given by the ratio of shallow water wave speed to the background rotation rate, $L_{R}=\sqrt{gD}/2\Omega_{a}$.
This dispersion relation predicts the maximum Rossby wave frequency, $\omega_{max}$, to occur for wavelengths approaching $L_{R}$, where,
\begin{equation}
\omega_{max} = \frac{|\beta{L_{R}}|}{2}.
\label{eqn:omegamax}
\end{equation}
One implication of this limit is that forcing the system with frequencies greater than $\omega_{max}$ cannot generate propagating Rossby waves.
We can also use the dispersion relation to show that the zonal phase speed $c$ of Rossby waves is,
\begin{equation}
c =\frac{\omega}{k} = \frac{-\beta{L_{R}^{2}}}{1+L_{R}^{2}\left(k^{2}+l^{2}\right)}.
\end{equation}
Here it is apparent that for positive $\beta$, the zonal phase speed is negative definite, such that the zonal direction of Rossby wave celerity is always westward.
In the reference frame of the rotating annulus, the magnitude of the effective $f$ (Eqn. \ref{eqn:effective_f}) decreases outwards as $d$ increases, such that the outer regions of the annulus are analogous to lower latitudes, the inner regions are analogous to higher latitudes, and $\beta$ is positive.
Thus, the equivalent eastward and westward directions are cyclonic (with $\Omega_{a}$) and anti-cyclonic (against $\Omega_{a}$), respectively, and Rossby waves will propagate zonally anti-cyclonically around the annulus.
For the limit of very long waves $(\lambda_{k}, \lambda_{l}>>L_{R})$, the zonal phase speed approaches a maximum value of,
\begin{equation}
c_{max} =-\beta{L_{R}^{2}}.
\label{eqn:speedmax}
\end{equation}

In the case of a background eastward zonal flow of speed $U_{0}$, the Rossby wave dispersion relation becomes,
\begin{equation}
\omega = U_{0}k - \frac{\beta{k}}{\left(k^{2}+l^{2}\right)}.
\label{eqn:doppler_dispersion}
\end{equation}
This state permits the special case of $\omega=0$, whereby the Rossby waves become standing or stationary waves with zero frequency, corresponding to,
\begin{equation}
U_{0}k = \frac{\beta{k}}{\left(k^{2}+l^{2}\right)},
\end{equation}
which has the solutions $k=0$ and,
\begin{equation}
U_{0} = \frac{\beta}{\left(k^{2}+l^{2}\right)}.
\end{equation}
We can define the total horizontal wave number squared as $K^{2}=k^{2}+l^{2}$, and rearrange for $K$ for,
\begin{equation}
K = \sqrt{\frac{\beta}{U_{0}}}.
\end{equation}
That is, the expected wavelength of stationary Rossby waves in a background flow $U_{0}$ is,
\begin{equation}
\lambda_{K} = 2\pi\sqrt{\frac{U_{0}}{\beta}}.
\label{eqn:pedlosky_pred}
\end{equation}

\subsection{Differential Sidewall Heating of a Rotating Annulus}

Now we consider the case that the inner and outer sidewalls of the rotating annulus are maintained at distinct temperatures $T_{i}$ and $T_{o}$, respectively, with a constant temperature difference $\Delta{T} = T_{o}-T_{i}$, such that the system is a Hide Tank \cite[e.g.,]{hide1958}.
Following on from the more complete introductions of Hide Tank dynamics by \cite{read2011} and \cite{read_etal2015}, and the recent reviews of \cite{stewart_shakespeare2024} and \cite{harlander_etal2024}, flows in Hide Tanks are primarily characterised by their thermal Rossby number $Ro_{T}$,
\begin{equation}
Ro_{T} = \frac{g\alpha|\Delta{T}|D}{\Omega_{a}^{2}L^{2}},
\label{eqn:thermalrossby}
\end{equation}
where $\alpha$ is the thermal expansion coefficient, and their Taylor number $Ta$,
\begin{equation}
Ta = \frac{4\Omega_{a}^{2}L^{5}}{\nu^{2}D},
\end{equation}
where $\nu$ is the kinematic viscosity.
The thermal Rossby number $Ro_{T}$ represents the ratio of the buoyancy force to the Coriolis force, and the Taylor number $Ta$ represents the ratio of the Coriolis force to the viscous force.
These parameters are particularly useful for describing the dynamical regime of a system, and comparing experiments from different apparatuses.
For example, systems with $Ro_{T}<O(10^{-1})$ and $Ta>O(10^{8})$ are in a regime of geostrophic turbulence.

The magnitude of the sidewall temperature difference $|\Delta{T}|$ allows for the development of a stratification that has an upper bound of,
\begin{equation}
N=\sqrt{g\alpha\frac{|\Delta{T}|}{D}}.
\end{equation}
In addition to maintaining the stratification, the differential sidewall thermal forcing will excite a bulk zonal geostrophic flow with velocity $u_{g}$ that scales according to the thermal wind equation,
\begin{equation}
u_{g} \propto \frac{g\alpha\Delta{T}D}{2\Omega_{a}{L}},
\end{equation}
where $L$ is the maximum radial lengthscale of the annulus, $L = r_{o}-r_{i}$.
In practice, these thermally-driven bulk geostrophic flows generally tend to be relatively slow, perhaps only a few millimetres per second.
That said, the flows are typically unstable and for certain conditions will develop into background geostrophic turbulence by way of baroclinic and lateral shear instabilities.

An upper limit for localised flow arising from the differential sidewall thermal forcing is the theoretical speed of a gravity current $u_{c}$ driven by the maximum temperature difference,
\begin{equation}
u_{c} = \frac{\sqrt{g\alpha|\Delta{T}|D}}{2}.
\end{equation}
This limit is useful to obtain a scaling for the baroclinic Rossby deformation radius $L_{D}$ as the ratio of the gravity current speed $u_{c}$ to the background rotation,
\begin{equation}
L_{D} = \frac{u_{c}}{2\Omega_{a}} = \frac{\sqrt{{g\alpha|\Delta{T}|D}}}{4\Omega_{a}}.
\label{eqn:baroclinicrossby}
\end{equation}
The background geostrophic turbulence tends to concentrate energy at the lengthscale of this baroclinic deformation radius.

The squared ratio of the baroclinic Rossby deformation radius to the width of the annulus provides another useful dimensionless parameter known as the Burger number $Bu$,
\begin{equation}
Bu = \left(\frac{L_{D}}{L}\right)^{2}.
\label{eqn:burger}
\end{equation}
In the regime $Bu<<1$, the domain is physically large enough for baroclinic geostrophic turbulence at the deformation scale $L_{D}$ to develop and evolve without boundary interference or wall effects.
Note the dynamical equivalence of the Burger number $Bu$ to the thermal Rossby number $Ro_{T}$; comparing Equations \ref{eqn:thermalrossby}, \ref{eqn:baroclinicrossby} \& \ref{eqn:burger} implies $Ro_{T}=16Bu$.

\subsection*{Sign of the Sidewall Temperature Difference}

The sidewall temperature difference maintains thermal structure throughout the interior of the fluid; in addition to the vertical stratification $N$, there is also a non-zero radial temperature gradient $\partial{T}/\partial{r}$.
The warmest fluid in the annulus will exist at the surface immediately adjacent to the warmer sidewall, and the coldest fluid in the annulus will be at the base near the cool sidewall.
Given the background rotation of the annulus, it is possible for all levels through the fluid depth to maintain a non-zero horizontal temperature gradient such that the isotherms between the warmest and coolest fluids are sloping.
That is, the isopycnals can maintain both vertical and radial structure in the time mean.
Now consider the fact that the sidewall temperature difference, defined as $\Delta{T}= T_{o} - T_{i}$, can be negative or positive.
This implies that the isopycnal surfaces can slope either up or down with radius (e.g., Fig. \ref{fig:01_LRA_schematic}d); for a negative sidewall temperature difference, the isopycnal surfaces tend to slope up with radius, and vice versa.

Recall that the free surface of a sufficiently rapidly rotating annulus slopes up with radius (Eqn. \ref{eqn:d_by_r_Omega}).
Given that the hydrostatic pressure of the fluid is simply a function of depth, the isobaric surfaces throughout the annulus follow the free surface and also slope up with radius.
Thus, a negative sidewall temperature difference (warmer at the inner sidewall) will have isopycnal and isobaric surfaces that both slope up with radius in a similar direction.
A positive sidewall temperature difference (warmer at the outer sidewall), however, will have isopycnal and isobaric surfaces with opposing slopes.
An important measure of the baroclinicity of a fluid is related to the relative angle between the isopycnal and isobaric surfaces (e.g., $\nabla{P}\times\nabla{\rho}$); this implies that the case of positive $\Delta{T}$ is expected to be relatively more baroclinic than the case of a negative $\Delta{T}$ with similar magnitude.

\section{Experiments}

\subsection{Laboratory Apparatus \& Methodology}

Our experiments here use the Large Rotating Annulus (LRA) facility in the Climate \& Fluid Physics Laboratory at the Australian National University (Fig. \ref{fig:01_LRA_schematic}a,b).
This unique facility is comprehensively described by \cite{stewart_shakespeare2024}; here we provide essential scientific detail and technical modifications specific to our experiments.
The LRA facility is a large 0.4\,m deep annular perspex annular tank with an inner and outer radii of $r_{i}=0.2$ and $r_{o}=0.8$\,m, respectively.
The annulus is filled with a working fluid of freshwater ($\rho_{0}=998$\,kg/m$^3$) to a reference depth of $D=0.25$\,m.
The annulus is mounted on a table that is able to be rotated about its central vertical axis; here we use 8 different clockwise rotation rates spanning from $\Omega_{a} = -0.4$ to $\Omega_{a} = -2.5$\,rad/s, inclusive, at intervals of $0.3$\,rad/s.
The inner and outer sidewall temperatures are able to be maintained at different constant temperatures, $T_{i}$ and $T_{o}$, respectively; here we focus on 4 distinct sidewall temperature configurations: ($T_{i},\,T_{o}$) $\approx$ (20,\,28), (21,\,23), (23,\,21), and (27,\,21)$^{\circ}$C.

One of the unique aspects of the LRA is a differentially-rotating outer ring to which bottom topography can be mounted and driven around the annulus.
The topography attachment is modular and able to accommodate a range of different 3D-printed pieces of bottom topography.
In our experiments here, we use a single 3D approximately-Gaussian bump that is $h_{b}=0.18$\,m high and with a half-width of $w_{b}=0.1$\,m.
This bump is located at mid-radius ($r_{b}=0.5$\,m) in the annulus and differentially-rotated at rates spanning between $\Omega_{b} = \pm0.11$\,rad/s, inclusive, at intervals of $0.02$\,rad/s; anti-clockwise relative to the annulus is positive $\Omega_{b}$.
These bump rotation rates correspond to bump velocities spanning the range of $U_{b}=r_{b}\Omega_{b} = \pm0.065$\,m/s.
Rotating the bump relative to the annulus generates flow-topography interactions; by analysing the experiments in the reference frame of the bump, the ambient fluid effectively circulates zonally around the annulus at a velocity dependent on $\Omega_{b}$.

The primary diagnostic used in the experiments here is a FLIR E75 thermal camera mounted approximately 3.2\,m above the water surface on a frame that co-rotates with the table.
The spatial resolution of the camera is at least $5\times5$\,mm$^2$, and covers approximately 95\% of the water surface area at any given time; the field of view of the thermal camera (42${^{\circ}}\times$32${^{\circ}}$) does not quite span the entire surface of the annulus in both horizontal directions, with approximately 10\,cm of the annulus out of view in the short direction.
Thick blackout curtains completely enshroud the LRA so as to minimise evaporation of the water, maintain the temperature and humidity of the air above the water to facilitate accurate infrared measurements, and to rotate this overlying ambient air with the annulus so as to eliminate surface wind stresses.
For each experiment, surface temperature data is collected from the thermally-equilibrated circulation states at 1\,Hz for 2 hours.
The time taken for the experiment to reach a state of thermal equilibrium depends on the initial condition of the apparatus and working fluid, and can be between 1 hour (for a minor change in bump rotation rate) and 24 hours (for a sidewall temperature adjustment and subsequent re-stratification process); during this time the approach to thermal equilibrium is monitored by a series of onboard thermistors logging the temperatures throughout the working fluid and overlying ambient air.

\subsection{Parameter Space}

\begin{figure}[]
\centering
\includegraphics[width=1.0\textwidth]{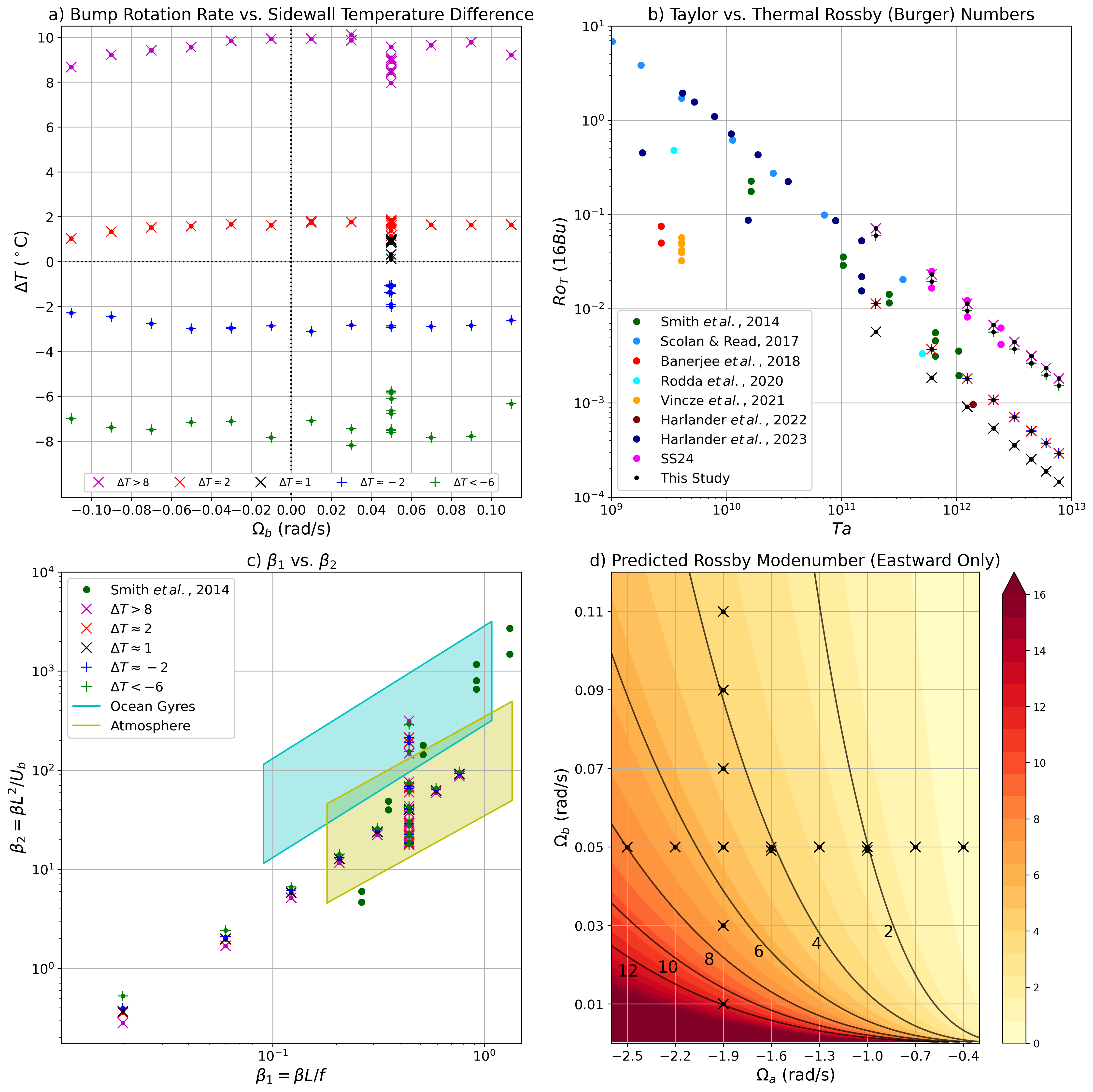}
\caption{The dynamical parameter space covered by the experiments. Panel (a) shows the sidewall temperature differences $\Delta{T}$ and topographic bump rotation rate $\Omega_{b}$ for the 93 different experiments. The Taylor and thermal Rossby numbers of our experiments are shown in (b), along with those of recent studies in Hide Tanks; our experiments here are indicated with $\times$ and $+$ symbols representing positive and negative sidewall temperature differences, respectively. Panel (c) shows the planetary and planetary-relative vorticity numbers, $\beta_{1}$ and $\beta_{2}$, respectively, for our experiments and those of \cite{smith_etal2014}; also shown are the regime spaces for mid-latitude atmospheric dynamics and ocean gyres. Panel (d) shows the predicted modenumber of standing Rossby waves as a function of the annulus and bump rotation rates, $\Omega_{a}$ and $\Omega_{b}$, respectively, and where our eastward ($\Omega_{b}>0$) experiments occur in this space.}
\label{fig:02_parameterspace}
\end{figure}

Here we report on a series of 93 experiments.
These experiments primarily explore the effects of three independently controllable variables; the annulus rotation rate $\Omega_{a}$, the bump rotation rate $\Omega_{b}$, and the sidewall temperature difference $\Delta{T}$ (Fig. \ref{fig:02_parameterspace}a).
These experiments can be broadly grouped into two sets: the first set explores the effect of the annulus rotation rate, and the second set explores the effect of the bump rotation rate, with both sets spanning a range of sidewall temperature differences $\Delta{T}$.
For experiments that focus on the effect of the annulus rotation rate $\Omega_{a}$, we examine 8 different rates that span $\Omega_{a}=-0.4$ to $\Omega_{a}=-2.5$\,rad/s at 0.3\,rad/s intervals; these experiments are all run with a constant bump rotation rate of $\Omega_{b}=0.05$\,rad/s.
For experiments that focus on the effect of the bump rotation rate $\Omega_{b}$, we use 12 different rates that span $\Omega_{b}=\pm0.11$\,rad/s at 0.02\,rad/s intervals; these experiments are all run with a constant annulus rotation rate of $\Omega_{a}=-1.9$\,rad/s.
In both sets of experiments, all 4 different sidewall temperature differences are used; this gives at least $8\times4$ experiments for the first set, and at least $12\times4$ experiments for the second set (several cases were duplicated to test experiment repeatability).

The experiments range in Taylor numbers between $Ta\approx2\times10^{11}$ and $Ta\approx8\times10^{12}$, thermal Rossby numbers between $Ro_{T}\approx1.5\times10^{-4}$ and $Ro_{T}\approx7\times10^{-2}$, and Burger numbers between $Bu\approx2.3\times10^{-3}$ and $Bu\approx1.1$ (Fig. \ref{fig:02_parameterspace}b).
These experiments cover and extend the dynamical range of cases presented by \cite{stewart_shakespeare2024} in the LRA, and are entirely within the regime of geostrophic turbulence ($Ta\gtrsim{10^{10}}$, $Ro_{T}\lesssim{10^{-1}}$).
Included for reference are recent experiments of \cite{smith_etal2014}, \cite{rodda_etal2020}, \cite{vincze_etal2021}, and \cite{harlander_etal2022} which all use traditional Hide Tanks with sidewall temperature differences, as well as the experiments of \cite{scolan_read2017}, \cite{banerjee_etal2018}, and \cite{harlander_etal2023} which use the ``Fultz'' style of thermal boundary conditions with heating at the lower periphery and cooling at the upper centre \cite[see, e.g.,][]{fultz_etal1959, harlander_etal2023, stewart_shakespeare2024}.

The dynamical importance of the background radial gradient in potential vorticity is shown in Figure \ref{fig:02_parameterspace}c as a comparison of the terms $\beta_{1}$ and $\beta_{2}$ (Eqns. \ref{eqn:beta1} \& \ref{eqn:beta2}, respectively).
For the large scale atmospheric dynamics at mid-latitudes, these terms span between $\beta_{1}\approx0.2$--$1.1$ and $\beta_{2}\approx5$--500; this is dynamically equivalent to the Coriolis parameter changing by at least 20\% in the region of interest, and that the change in relative vorticity in the region of interest is at least 5 times smaller than the change of the Coriolis parameter.
For large scale ocean gyres, these bounds are understandably different at $\beta_{1}\approx0.1$--$1.0$ and $\beta_{2}\approx10$--2000.
The experiments presented here span the range of $\beta_{1}\approx0.05$--$1.1$ and $\beta_{2}\approx0.5$--400; many of these cases exist in the dynamical regime equivalent to that of large scale mid-latitude atmospheric dynamics.
This is in contrast with the similar experiments of \cite{smith_etal2014}, who focussed their investigation on large scale ocean dynamics.

The bump has a prescribed zonal velocity of $U_{b}=r_{b}\Omega_{b}$, which is dynamically equivalent to a background zonal flow speed when the system is analysed in the frame of reference of stationary bump.
The combination of a zonal flow and background gradient in potential vorticity $\beta$ allow us to calculate the expected wavelength of standing Rossby waves by Equation \ref{eqn:pedlosky_pred}, which can subsequently be converted into a modenumber by considering the circumference of the annulus at the radius of the bump, $2{\pi}r_{b}$.
Figure \ref{fig:02_parameterspace}d shows the predicted modenumber of standing Rossby waves for the eastward flowing cases (i.e., $U_{b}>0$).
For experiments that cover different annulus rotation rates, the predicted modenumbers of standing Rossby waves span from less than modenumber 2 for the lowest annulus rotation rates, and up to modenumber 8 for the highest annulus rotation rate.
Experiments that vary the bump rotation rate are predicted to have standing Rossby waves between modenumber 12 for the slowest bump rotation and approximately modenumber 3 for the fastest bump rotation.

\begin{table}
\begin{center}
\caption{Control and dimensionless parameters.}
\begin{tabular}{ccc}
Symbol & Description & Value/Range/Units \\
\hline
$\rho_{0}$ & Reference Density & 998\,kg/m$^{3}$ \\
$r_{i}$ & Inner Edge Wall Radius & 0.2\,m  \\
$r_{o}$ & Outer Edge Wall Radius & 0.8\,m  \\
$L$ & Annulus Width, $L = r_{o}-r_{i}$ & 0.6\,m  \\
$D$ & Reference Fluid Depth & 0.25\,m  \\
$\Omega_{a}$ & Annulus Rotation Rate & (-0.4, -2.5)\,rad/s \\
$d$ & Fluid Depth, $d\sim{f(r,\Omega_{a})}$ & m \\
$\nu$ & Kinematic Viscosity & 10$^{-6}$\,m$^{2}$/s  \\
$g$ & Gravity & 9.81\,m$^2$/s \\
\hline
$r_{b}$ & Annulus Radius at Bump & 0.5\,m  \\
$\Omega_{b}$ & Bump Rotation Rate & (-0.11, 0.11)\,rad/s \\
$U_{b}$ & Bump Velocity, $U_{b}=r_{b}\Omega_{b}$ & (-0.055, 0.055)\,m/s \\
\hline
$T_{i}$ & Inner Edge Wall Temperature & (20, 27)$^\circ$C  \\
$T_{o}$ & Outer Edge Wall Temperature & (21, 28)$^\circ$C  \\
$\Delta{T}$ & Sidewall Temperature Difference & (-6, 8)$^\circ$C  \\
$N$ & Buoyancy Frequency & (0.03,0.28)\,rad/s \\
\hline
$Ro$ & Rossby Number & $7\times{10^{-4}}<Ro<2\times{10^{-2}}$  \\
$Ro_{T}$ & Thermal Rossby Number & $1.4\times{10^{-4}}<Ro_{T}<7\times{10^{-2}}$    \\
$Ta$ & Taylor Number & $2\times{10^{11}}<Ta<8\times{10^{12}}$    \\
$Bu$ & Burger Number & $2.3\times{10^{-3}}<Bu<1.1$ \\
$f$ & Coriolis Parameter, $f = 2\Omega_{a}$ & (-0.8, -5.0)\,rad/s \\
$\zeta$ & Relative Vorticity, $\zeta=\partial{v}/\partial{x}-\partial{u}/\partial{y}$ & s$^{-1}$ \\
$q$ & Potential Vorticity, $q=(f+\zeta)/d$ & s$^{-1}$ \\
$\beta$ & Radial Gradient of Potential Vorticity & $0.026<\beta<6.37$    \\
$\beta_{1}$ & Planetary Number & $ 0.04<\beta_{1}<1.2$  \\
$\beta_{2}$ & Planetary-Relative Vorticity Number & $0.3<\beta_{2}<400$    \\
\hline
$L_{R}$ & Barotropic Rossby Deformation Radius & (0.31, 1.9)\,m\\
$L_{D}$ & Baroclinic Rossby Deformation Radius & (0.01, 0.52)\,m\\
$\omega_{max}$ & Rossby Wave Maximum Frequency & (0.02, 0.99)\,s$^{-1}$ \\
$c_{max}$ & Rossby Wave Maximum Phase Speed & (0.1, 0.63)\,m/s \\
$\lambda_{K}$ & Predicted Wavelength of Standing Rossby Wave & (0.26, 1.57)\,m \\
\hline
\end{tabular}
\label{tab: exp_parameters}
\end{center}
\end{table}

\subsection{Analysis}

\begin{figure}[]
\centering
\includegraphics[width=1.0\textwidth]{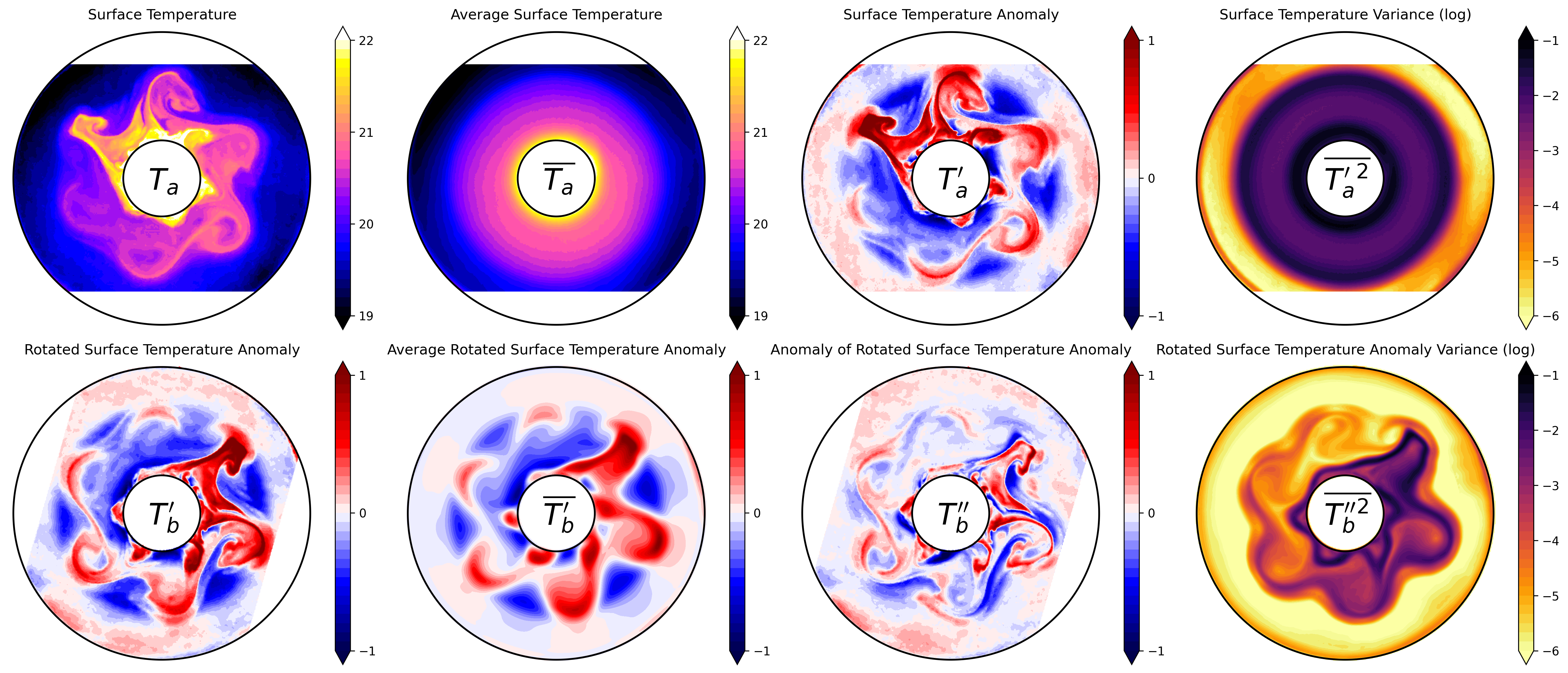}
\caption{A schematic representation of the analysis process to go from the snapshots of surface temperatures $T_{a}$, taking the time mean $\overline{T_{a}}$, and calculating the time-dependent anomaly $T^{\prime}_{a}$, where $T_{a}(t) = \overline{T_{a}} - T^{\prime}_{a}(t)$ for some time $t$, and variance $\overline{{T_{a}^{\prime}}^{2}}$. The anomaly $T^{\prime}_{a}$ is then rotated frame-by-frame about the centre of the image by a angle that depends on the bump rotation rate $\Omega_{b}$ such that the bump remains stationary between successive frames; this rotated anomaly is referred to as $T^{\prime}_{b}$. The time mean $\overline{T^{\prime}_{b}}$ is then taken and used to calculate the time-dependent anomalies of the rotated temperature anomaly field $T^{\prime\prime}_{b}$, where $T^{\prime}_{b}(t) = \overline{T^{\prime}_{b}} - T^{\prime\prime}_{b}(t)$, and variance $\overline{{T_{b}^{\prime\prime}}^{2}}$.}
\label{fig:03_analysis_workflow}
\end{figure}

The primary diagnostic for the experiments presented here are 2-hour timeseries of equilibrated surface temperatures obtained from a thermal camera rotating with the annulus and recording at 1\,Hz.
These 7200 snapshots represent surface temperatures captured in the frame of reference of the annulus $T_{a}$ (where the subscript $a$ denotes the annulus frame of reference), such that the bump appears to circumnavigate the annulus in time through the background geostrophic turbulence (Fig. \ref{fig:03_analysis_workflow}).
We first calculate the time-average of surface temperatures in the annulus reference frame $\overline{T_{a}}$ (where the overline denotes a time-mean over an integer number of bump laps).
These time-average fields resemble concentric circles of isotherms and are indicative of the background radial temperature gradient; the azimuthal location of the bump is averaged out over many laps through time.
Using these time-average fields we can then calculate the time-dependent temperature anomalies $T^{\prime}_{a}$, again in the frame of reference of the rotating annulus, where $T^{\prime}_{a} = \overline{T_{a}} - T_{a}$.
The thermal structure of the time-dependent temperature anomalies appear consistent with the original snapshots, but have the background radial gradient in temperature removed.
The time-average of these anomalies squared returns the surface temperature variance, $\overline{{T_{a}^{\prime}}^{2}}$.

The time-dependent anomaly fields $T^{\prime}_{a}$ are then rotated frame-by-frame about the pixel that contains the centre of the annulus and at the rotation rate of the bump; this procedure returns a timeseries of rotated time-dependent temperature anomalies $T^{\prime}_{b}$ (where the subscript $b$ denotes the bump frame of reference) wherein the bump is stationary, with the annulus and fluid appearing to rotate and flow relative to this bump, respectively.
The time-average of these rotated time-dependent temperature anomaly fields is then taken $\overline{T^{\prime}_{b}}$.
These time-average rotated anomaly fields preserve features that are standing or stationary relative to the bump; the process of averaging through 7200 frames ensures that any structure that survives this analysis step is indeed a robust and persistent feature of the system.
The anomaly of these rotated time-dependent temperature anomaly fields is also taken, $T^{\prime\prime}_{b}$; these double anomaly fields quantify the variability associated with the bump-relative non-stationary or transient components of the flow.
Again, any structure that appears in these rotated double anomaly fields is indeed a robust and persistent feature of the system.
The time-average of these double anomalies squared returns the rotated surface temperature anomaly variance, $\overline{{T_{b}^{\prime\prime}}^{2}}$.

Separating the rotated time-dependent temperature anomalies $T^{\prime}_{b}$ into their time-mean $\overline{T^{\prime}_{b}}$ and time-dependent $T^{\prime\prime}_{b}$ components allows us to compare the total variability contained within these partitions of the flow.
We use the expression relating these terms, square it, and take its time- and spatial means;
\begin{equation}
T^{\prime}_{b} = \overline{T^{\prime}_{b}} - T^{\prime\prime}_{b}\quad \rightarrow \quad{T^{\prime}_{b}}^{2} = {\overline{T^{\prime}_{b}}}^{2} -2\overline{T^{\prime}_{b}}T^{\prime\prime}_{b}+ {T^{\prime\prime}_{b}}^{2}\quad \rightarrow \quad \overline{\langle{T^{\prime}_{b}}^{2}\rangle} = {\langle{\overline{T^{\prime}_{b}}}^{2}\rangle} + \overline{\langle{T^{\prime\prime}_{b}}^{2}\rangle},
\label{eqn:energy_terms}
\end{equation}
where the angled brackets denote the spatial means.
The terms $\overline{\langle{T^{\prime}_{b}}^{2}\rangle}$, ${\langle{\overline{T^{\prime}_{b}}}^{2}\rangle}$, and $\overline{\langle{T^{\prime\prime}_{b}}^{2}\rangle}$ represent the total thermal variability, the thermal variability contained in the standing features, and the thermal variability contained in the transient features, respectively.
Comparing the relative sizes of these total, standing and transient variability terms provides an indication of the energy partitioned within these aspects of the flow.

\section{Results \& Discussions}

We present the results in three sections.
The first and second sections show qualitatively the effects of the annulus rotation rate and bump rotation rate, respectively; the analysis in these first two sections focus on the distributions of surface temperature.
The third section presents more quantitative analysis of the Rossby Waves and thermal variabilities for both experiment sets.

\subsection{Qualitative Effect of Annulus Rotation Rate on Surface Temperatures}

\begin{figure}
\centering
\includegraphics[width=1.0\textwidth]{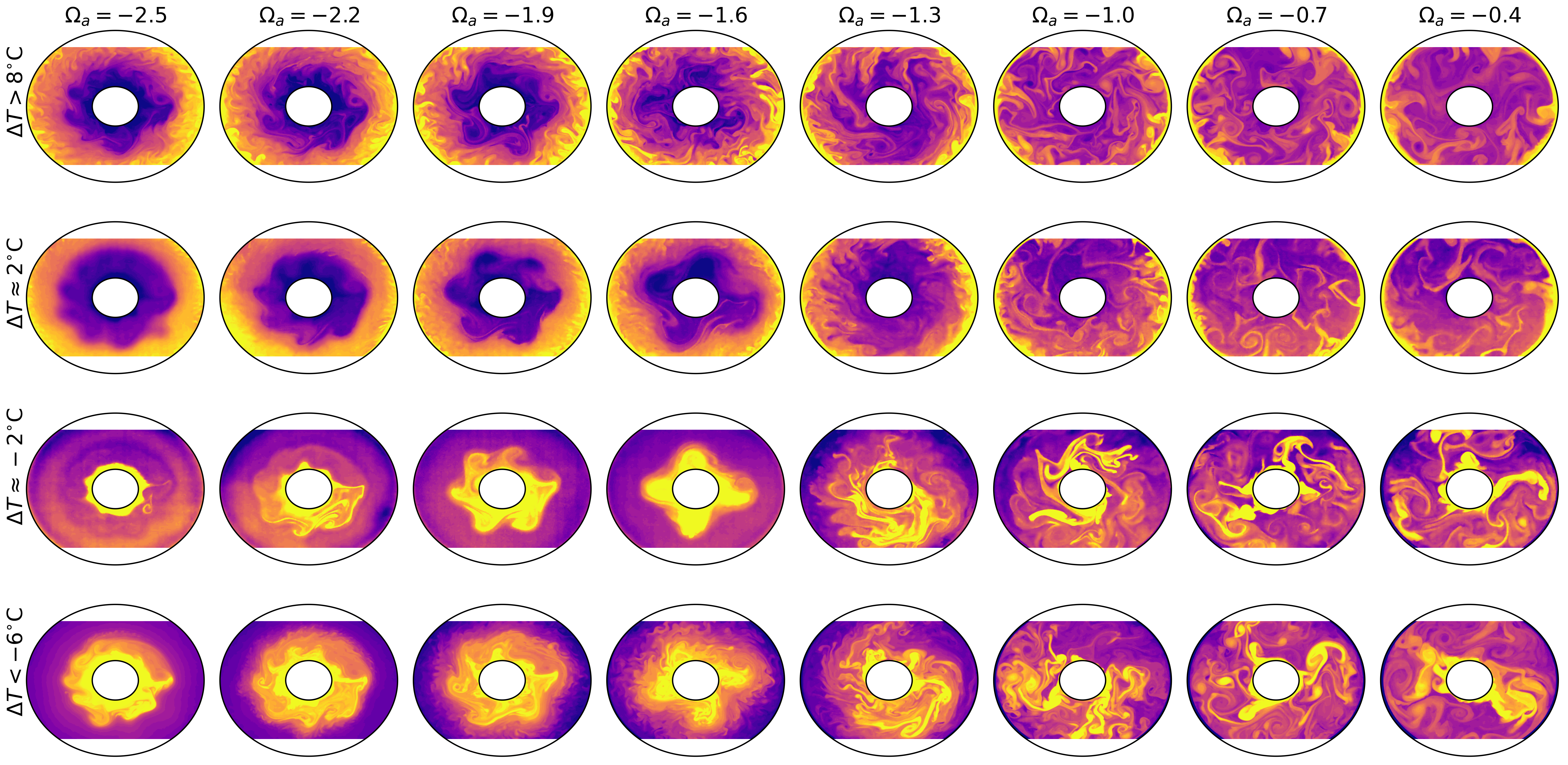}
\caption{Snapshots of the surface temperature fields $T_{a}$ for the cases with $\Delta{T}>8^{\circ}$C, $\Delta{T}\approx2^{\circ}$C, $\Delta{T}\approx-2^{\circ}$C, and $\Delta{T}<-6^{\circ}$C (top to bottom rows, respectively). The magnitude of the annulus rotation rate decreases from $\Omega_{a}=-2.5$\,rad/s at the left to $\Omega_{a}=-0.4$\,rad/s at the right in 0.3\,rad/s intervals. The colourscale in each case is centered about the frame mean temperature and spans 3 times the standard deviation; the intention of this normalised colourscale is to show the thermal structures of the flows rather than actual temperatures. The timing of the snapshot is selected such that the bump is located at approximately 3 o'clock. In all cases the bump rotation rate is $\Omega_{b}=0.05$\,rad/s, such that the background flow relative to the bump is clockwise.}
\label{fig:04_omegaa_sst_4x8}
\end{figure}

Snapshots of the surface temperature fields provide a useful means of assessing the characteristic nature of the flows.
Figure \ref{fig:04_omegaa_sst_4x8} shows surface temperature $T_{a}$ snapshots for all of the 8 different annulus rotation rates $\Omega_{a}$, and for 4 sidewall temperature differences: $\Delta{T}>8^{\circ}$C, $\Delta{T}\approx\pm2^{\circ}$C, and $\Delta{T}<-6^{\circ}$C.
The colourscale in each case is centered about the frame mean temperature and spans 3 times the standard deviation; this normalisation is done to highlight the spatial structures of the temperature fields rather than the absolute values of temperature.
All these cases have a constant bump rotation rate of $\Omega_{b}=0.05$\,rad/s; that is, the background flow relative to the bump is eastward (clockwise).
The cases with annulus rotation rates of $\Omega_{a}\leq|1.0|$\,rad/s (rightmost three columns) all exhibit filamented and eddying thermal structures consistent with geostrophic turbulence.
The lengthscales of the thermal features decreases as the magnitude of the annulus rotation rate increases from $\Omega_{a}=-0.4$ to $\Omega_{a}=-1.0$\,rad/s.
The thermal structure of the system appears to transition between $\Omega_{a}=-1.3$ and $\Omega_{a}=-1.6$\,rad/s; the latter cases exhibit structures of larger lengthscales and regular zonal wave-like features with zonal modenumber 4.
The modenumber of these zonal features increases as the magnitude of $\Omega_{a}$ increases; each 0.3\,rad/s increase in $|\Omega_{a}|$ appears to result in a zonal modenumber increase of approximately 1--2.
This increase in zonal modenumber is consistent with that predicted by the wavelength of standing Rossby waves (e.g., Eqn. \ref{eqn:pedlosky_pred}; Fig. \ref{fig:02_parameterspace}d).
The cases with the larger magnitude sidewall temperature difference (top and bottom rows) all exhibit relatively more small-scale structures, especially near to their respective warm sidewall.

\begin{figure}
\centering
\includegraphics[width=1.0\textwidth]{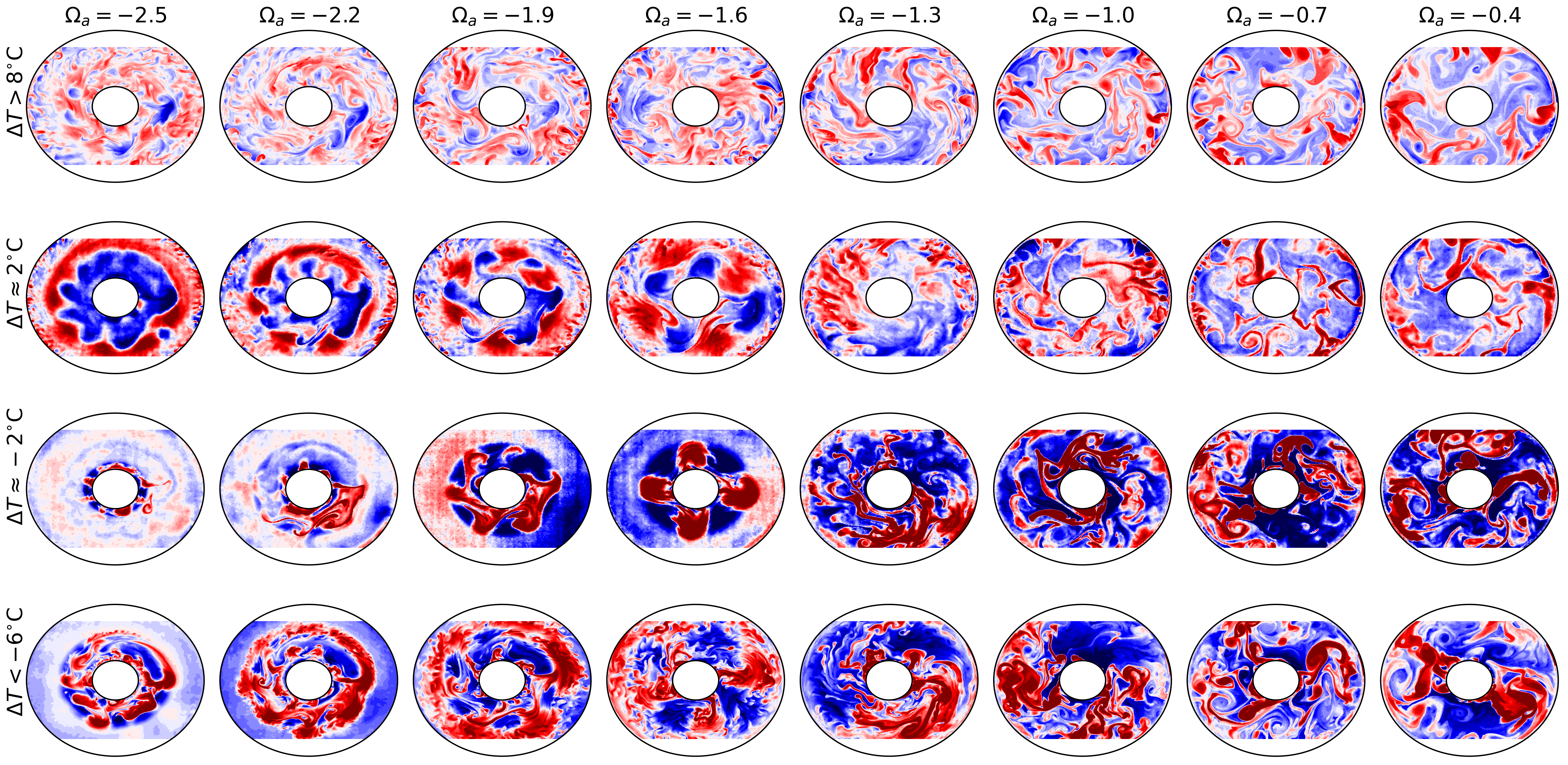}
\caption{Snapshots of the surface temperature anomaly fields $T^{\prime}_{a}$ for the same frames shown in Figure \ref{fig:04_omegaa_sst_4x8}. The colourscale in each case is normalised by half of the experiment sidewall temperature difference.}
\label{fig:05_omegaa_sst_anom_4x8}
\end{figure}

The structures of the surface temperature fields can be enhanced by removing the time-average surface temperature $\overline{T_{a}}$ to obtain snapshots of the surface temperature anomalies $T^{\prime}_{a}$ (Fig. \ref{fig:05_omegaa_sst_anom_4x8}).
These snapshots are the same frames as those of Figure \ref{fig:04_omegaa_sst_4x8}.
The filamented structures of the slower annulus rotation rate cases ($\Omega_{a}\leq|1.0|$\,rad/s) are more obvious, and tending to be associated with warm anomalies.
In the case of $\Omega_{a}=-1.3$\,rad/s, the thermal structures are appear to be tending to organise towards a zonal modenumber 1 distribution.
The faster annulus rotation rate cases ($\Omega_{a}\geq|1.6|$\,rad/s) show the relative signs of the thermal anomalies associated with the larger scale zonal waves; the ``peaks'' of the waves are associated with cool anomalies with an apparent cyclonic flow.
The comparison of the larger and smaller sidewall temperature difference cases shows remarkable similarity in the large scale structures, especially the zonal wave features.

\begin{figure}
\centering
\includegraphics[width=1.0\textwidth]{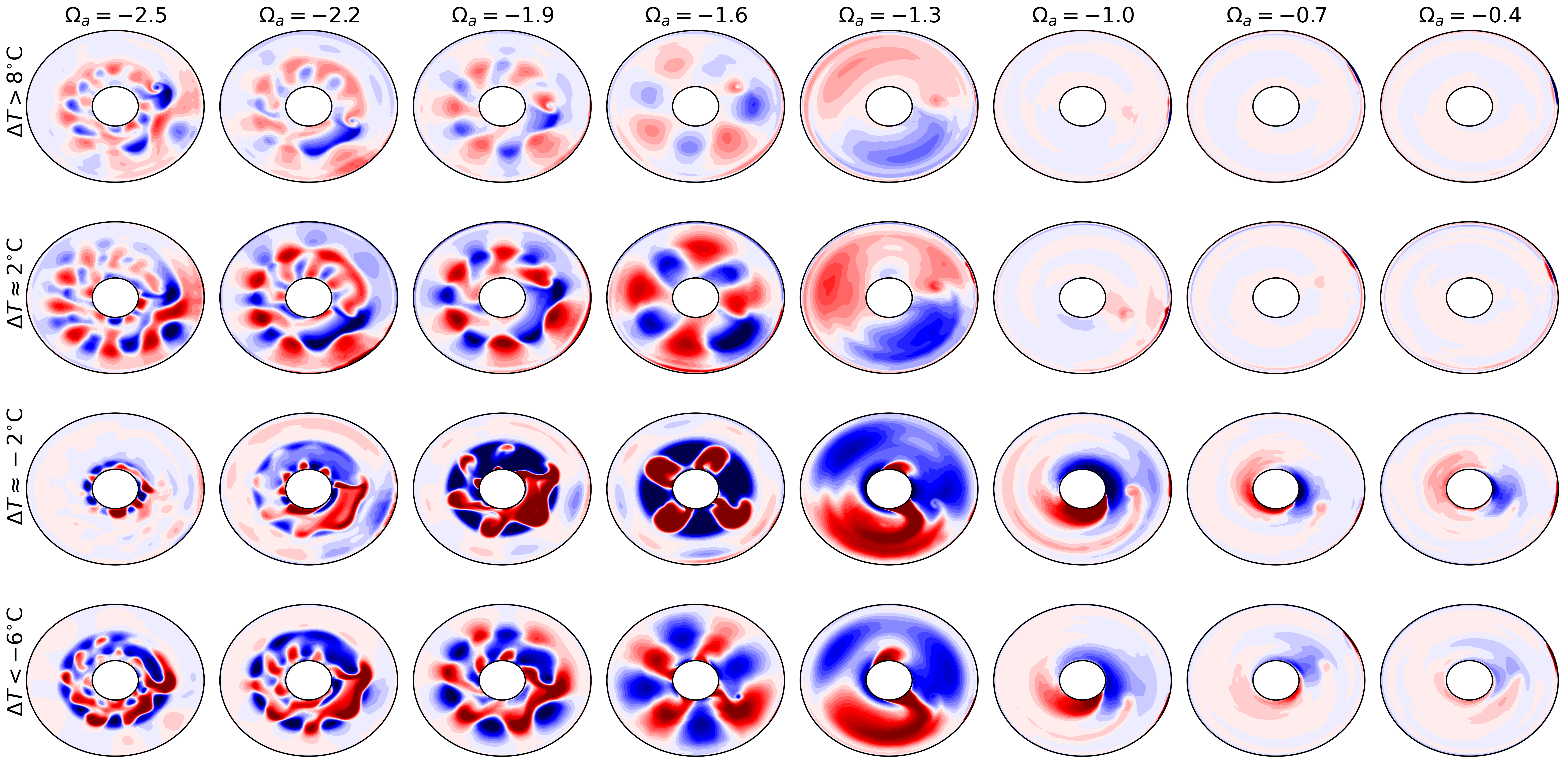}
\caption{Time-averages of the transformed surface temperature anomaly fields $\overline{T^{\prime}_{b}}$ for the same cases shown in Figures \ref{fig:04_omegaa_sst_4x8} \& \ref{fig:05_omegaa_sst_anom_4x8}. Again, the colour scales is normalised by half of the experiment sidewall temperature difference.}
\label{fig:06_omegaa_sst_anom_means_4x8}
\end{figure}

Transforming the surface temperature anomalies into the reference frame of the bump and taking its time-average returns the term $\overline{T^{\prime}_{b}}$ (Fig. \ref{fig:06_omegaa_sst_anom_means_4x8}); these fields show the thermal anomaly structures that are stationary relative to the bump.
The slower annulus rotation rate cases ($|\Omega_{a}|\leq1.0$\,rad/s) exhibit no substantial thermal anomaly structures that are standing relative to the bump; that is, there are no thermal anomalies that survive the time-averaging process in the frame of reference of the bump.
The case of $\Omega_{a}=-1.3$\,rad/s exhibits a zonal modenumber 1 distribution; cases with faster annulus rotation rates exhibit zonal waves with increasing modenumbers, consistent with that predicted in Eqn. \ref{eqn:pedlosky_pred} (Fig. \ref{fig:02_parameterspace}d).
The radial structure of the standing features appears to reduce for increasing $|\Omega_{a}|$.
The fastest annulus rotation rate cases ($|\Omega_{a}|\geq2.2$\,rad/s) appear to show some radial wave-like structure in addition to the standing zonal structures; these appear to be approximately similar radial and zonal wavelengths.

\subsection{Qualitative Effect of Bump Rotation Rate on Surface Temperatures}

Here we present the same series of analysis for a set of experiments exploring the effect of bump rotation rate $\Omega_{b}$.
Here we focus on all 12 settings of $\Omega_{b}$ (spanning $\pm0.11$\,rad/s at 0.02\,rad/s intervals), and 4 different sidewall temperature differences ($\Delta{T}>8^{\circ}$C, $\Delta{T}\approx2^{\circ}$C, $\Delta{T}\approx-2^{\circ}$C, $\Delta{T}<-6^{\circ}$C); all cases here have a constant annulus rotation rate of $\Omega_{a}=-1.9$\,rad/s, which is suitable for supporting standing zonal wave structures (third columns of Figs. \ref{fig:04_omegaa_sst_4x8}--\ref{fig:06_omegaa_sst_anom_means_4x8}).

\begin{figure}
\centering
\includegraphics[width=1.0\textwidth]{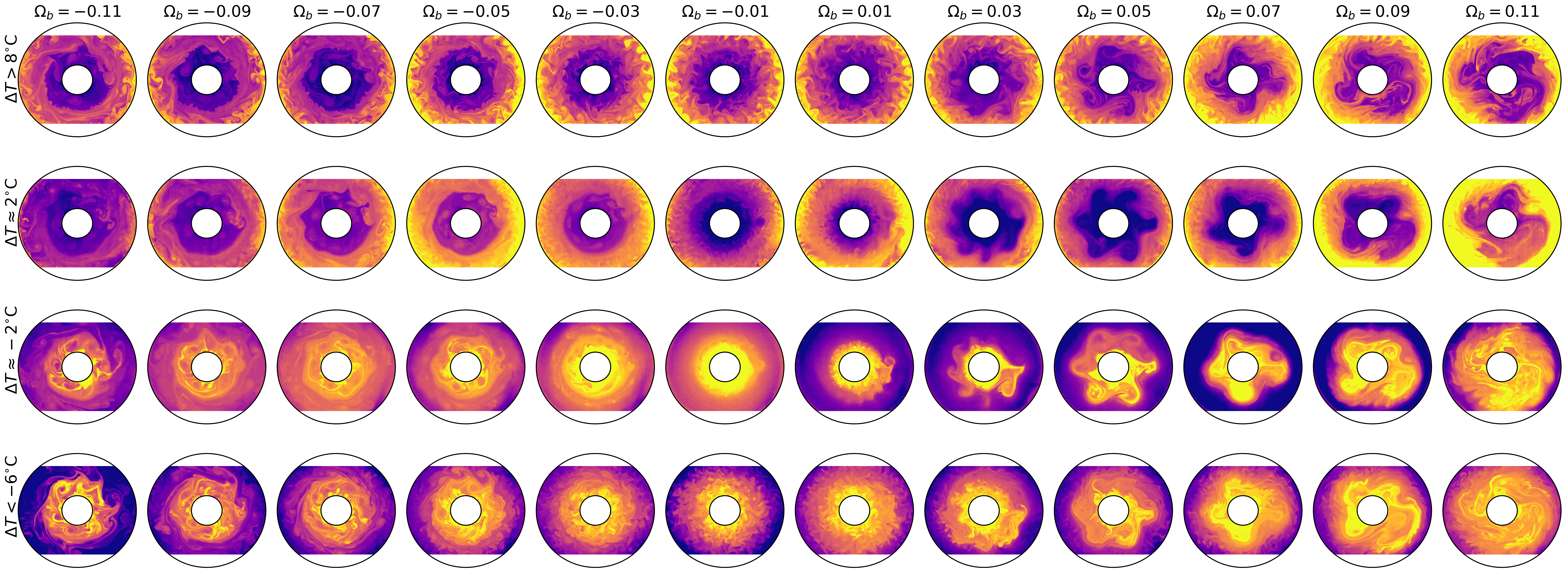}
\caption{Snapshots of the surface temperature fields $T_{a}$ for the cases with $\Delta{T}>8^{\circ}$C, $\Delta{T}\approx2^{\circ}$C, $\Delta{T}\approx-2^{\circ}$C, and $\Delta{T}<-6^{\circ}$C (top to bottom rows, respectively). The bump rotation rate increases from $\Omega_{b}=-0.11$\,rad/s at the left through 0 to $\Omega_{b}=0.11$\,rad/s at the right in 0.02\,rad/s intervals. The colour scale in each case is centered about the frame mean temperature and spans 3 times the standard deviation; the intention of this normalised colourscale is to show the thermal structures of the flows rather than actual temperatures. The timing of the snapshot is selected such that the bump is located at approximately 3 o'clock; the bump is travelling clockwise for cases with $\Omega_{b}<0$ (left half), and anti-clockwise for $\Omega_{b}>0$ (right half). In all cases the annulus rotation rate is $\Omega_{a}=-1.9$\,rad/s.}
\label{fig:07_omegab_snapshots_4x12}
\end{figure}

The snapshots of surface temperature highlight the richness of thermal structures present in this set of experiments (Fig. \ref{fig:07_omegab_snapshots_4x12}).
All experiments exhibit features consistent with strong geostrophic turbulence, especially along their respective warm sidewalls.
Cases with faster bump rotation rates ($|\Omega_{b}|\geq0.07$\,rad/s) exhibit filaments which appear to lengthen with $|\Omega_{b}|$.
Larger scale zonal waves are evident in the cases with eastward background flow relative to the bump ($\Omega_{b}>0$; rightmost 6 columns).
By contrast, the westward background flow cases ($\Omega_{b}<0$; leftmost 6 columns) do not have dominant zonal wave structure; however, these cases have a distinct strong radial gradient of temperature at mid-annulus (i.e. the radius of the bump), and some intermediate cases ($-0.09<\Omega_{b}<-0.05$\,rad/s) exhibit hexagonal circumpolar thermal structure around the annulus.
Note that the hexagonal circumpolar structures are reminiscent of Saturn's six-sided polar jet stream \cite[e.g.,][]{aguiar_etal2010, ingersoll2020, constantin2023}.

\begin{figure}
\centering
\includegraphics[width=1.0\textwidth]{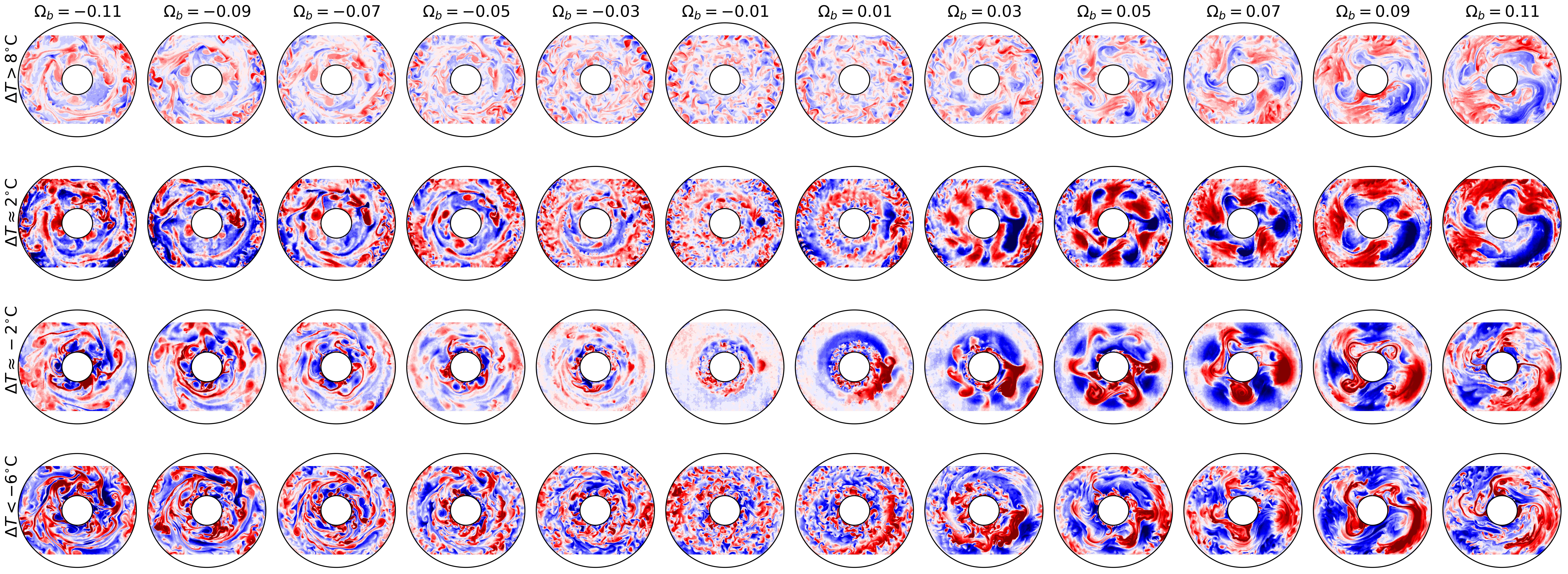}
\caption{Snapshots of the surface temperature anomaly fields $T^{\prime}_{a}$ for the same frames shown in Figure \ref{fig:07_omegab_snapshots_4x12}. The colourscale in each case is normalised by half of the experiment sidewall temperature difference.}
\label{fig:08_omegab_anom_snapshots_4x12}
\end{figure}

Snapshots of the surface temperature anomaly fields $T^{\prime}_{a}$ are shown in Figure \ref{fig:08_omegab_anom_snapshots_4x12}.
These highlight the effect that the flow direction can have on the characteristics of the thermal structures; eastward background flow cases ($\Omega_{b}>0$; rightmost 6 columns) exhibit larger scale zonal structures, while westward background flow cases ($\Omega_{b}<0$; leftmost 6 columns) tend to have filamented features and smaller scale eddies.
Experiments with a given bump rotation rate $\Omega_{b}$ all exhibit remarkable similar thermal structures for all 4 sidewall temperature differences.
The two sets of experiments with small sidewall temperature differences ($|\Delta{T}|\approx2^{\circ}$C; middle 2 rows) have normalised temperature anomalies of similar magnitudes, but with opposite sign anomalies.
By contrast, the two sets of experiments with large sidewall temperature differences ($|\Delta{T}|>6^{\circ}$C; top and bottom rows) exhibit different normalised temperature anomaly magnitudes; the cases with negative $\Delta{T}$ (bottom row) show normalised anomaly temperatures of similar magnitudes to the small temperature difference experiments, while the positive $\Delta{T}$ cases are notably weaker.
Recall that the isopycnal slopes of the large positive $\Delta{T}$ cases are opposed to the slopes of the isobars (Fig. \ref{fig:01_LRA_schematic}d); that is, the large positive $\Delta{T}$ cases are expected to be more prone to baroclinic instability relative to the large negative $\Delta{T}$ cases, and thus be relatively more mixed and with a narrower range of temperatures and reduced thermal variability signal.

\begin{figure}
\centering
\includegraphics[width=1.0\textwidth]{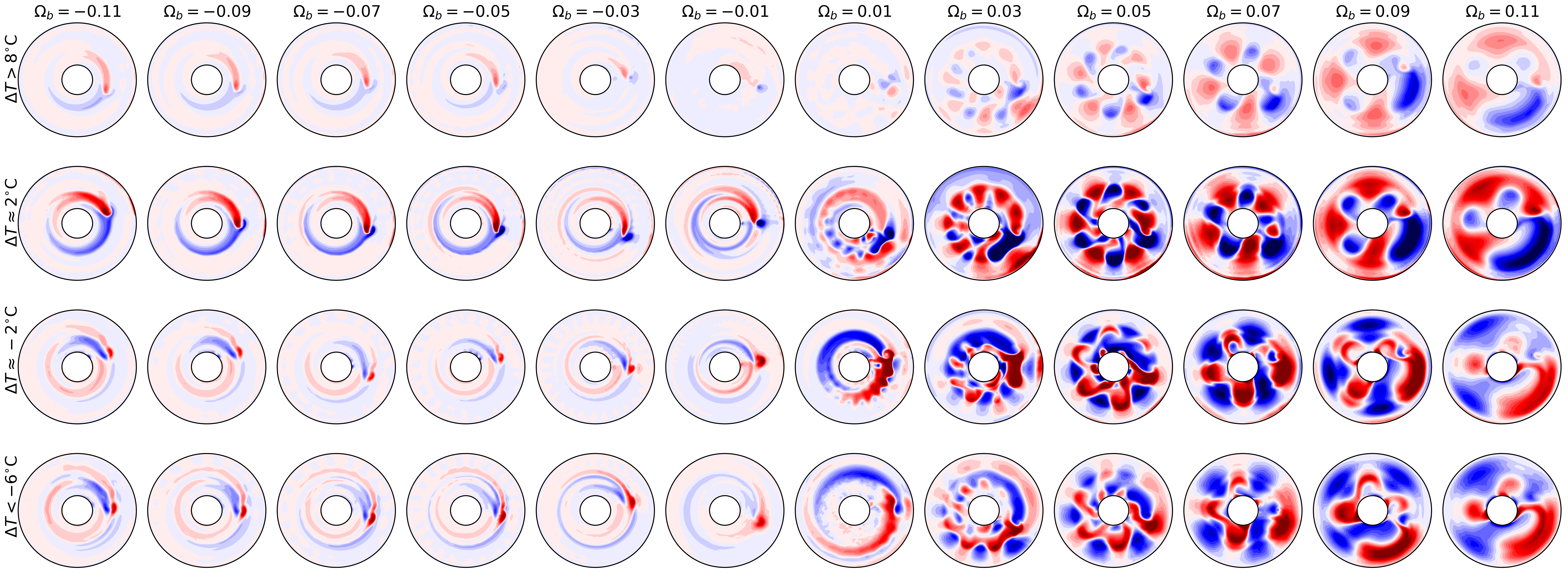}
\caption{Time-averages of the transformed surface temperature anomaly fields $\overline{T^{\prime}_{b}}$ for the same cases shown in Figures \ref{fig:07_omegab_snapshots_4x12} \& \ref{fig:08_omegab_anom_snapshots_4x12}. Again, the colour scales is normalised by half of the experiment sidewall temperature difference.}
\label{fig:09_omegab_anom_means_4x12}
\end{figure}

The transformed surface temperature anomaly means $\overline{T^{\prime}_{b}}$ shows the thermal anomaly structures that are stationary relative to the bump (Fig. \ref{fig:09_omegab_anom_means_4x12}).
Cases with westward flow ($\Omega_{b}<0$; leftmost 6 columns) exhibit mean thermal anomaly structure localised to the radius of the bump, with a maxima at the bump; the hexagonal circumpolar structures observed in the individual snapshots are not present in the time-average fields (i.e. these structures are not stationary relative to the topography). 
By contrast, the cases with eastward flow ($\Omega_{b}>0$; rightmost 6 columns) all exhibit substantial structure throughout the annulus and with zonal wave-like features.
The modenumber of these zonal waves appears to decrease with increasing bump speed.
Again, the structures have remarkably similar distributions across the different sidewall temperature differences; the smaller $|\Delta{T}|$ cases appear to just have opposing signed anomalies, while the larger $|\Delta{T}|$ cases exhibit a difference in normalised anomaly magnitudes with the positive $\Delta{T}$ cases notably relatively weaker.
The zonal structure of the cases with faster eastward flows ($\Omega_{b}>0.07$; rightmost 3 columns) appear to be less regular and somewhat deteriorated compared to the slower eastward flow cases.

\subsection{Quantitative Analysis}

\begin{figure}
\centering
\includegraphics[width=0.6\textwidth]{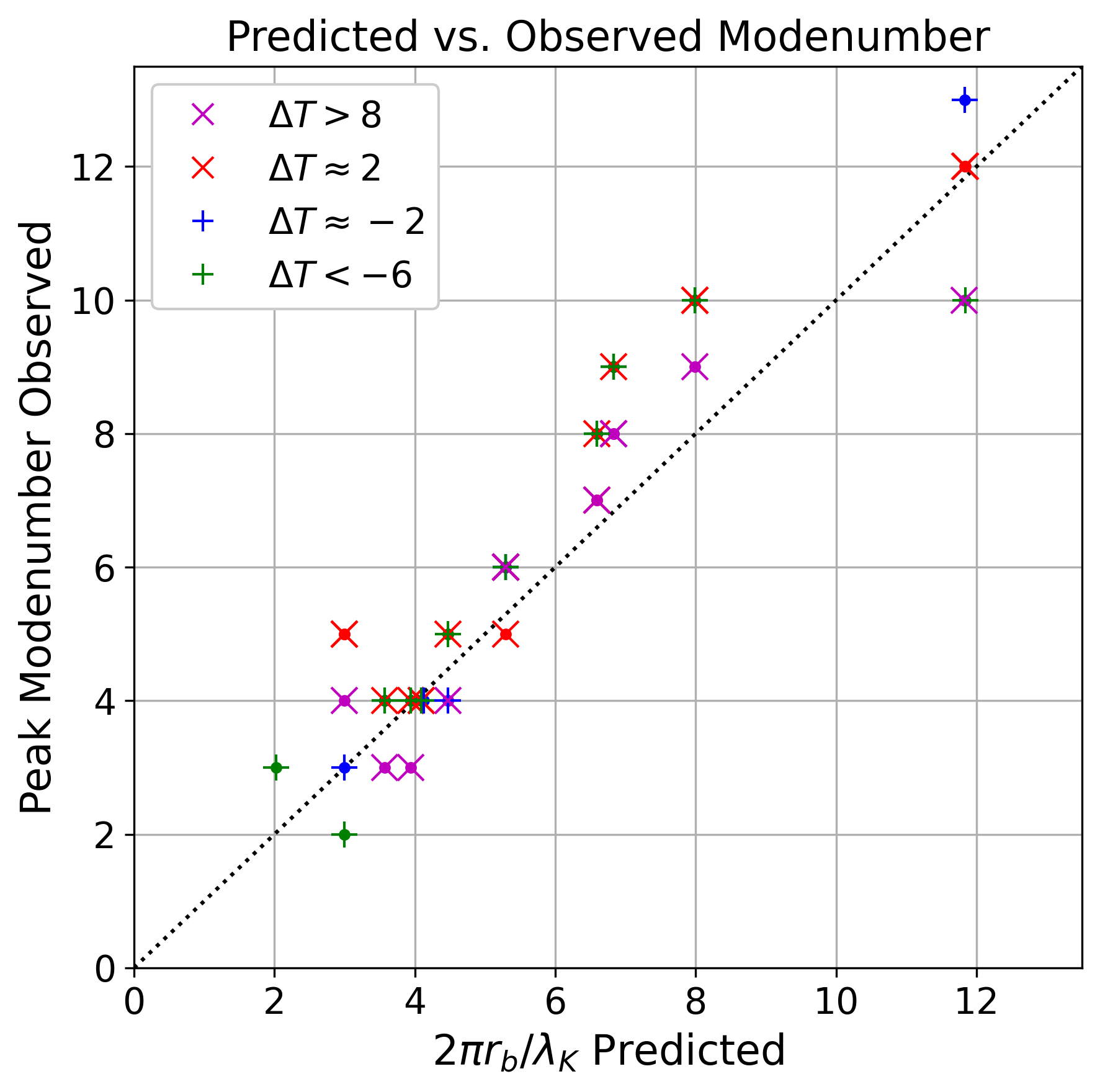}
\caption{The predicted modenumber of standing Rossby waves versus the observed peak modenumber for experiments with eastward background flow ($\Omega_{b}>0$).}
\label{fig:10_pedlosky}
\end{figure}

It is apparent that experiments with relatively rapid annulus rotation and eastward background flow exhibit zonal features that are stationary relative to the topography and reminiscent of standing Rossby waves.
Figure \ref{fig:10_pedlosky} shows a comparison of the observed dominant zonal modenumber (i.e. the zonal modenumber with maximum energy) to that predicted by Equation \ref{eqn:pedlosky_pred}; here the predicted modenumber is calculated by dividing the circumference of the annulus at the radius of the bump by the predicted wavelength, $2\pi{r_{b}}/\lambda_{K}$.
Note that the predicted wavelength is undefined for cases with westward flow relative to the bump; these cases are omitted from the analysis here.
Additionally, note that the observed dominant modenumber needs to be integer value, which is not required for the predicted modenumbers; this requirement is perhaps the largest source of uncertainty in this type of analysis.
That said, there is excellent agreement between the predicted and observed modenumbers; the range of this agreement extends between modenumber 2 up to approximately modenumber 12.
All the different sidewall temperature difference cases exhibit this agreement without an obvious trend; that is, there is no obvious difference between the negative and positive $\Delta{T}$ cases, or the large and small $|\Delta{T}|$ cases.

\begin{figure}
\centering
\includegraphics[width=1.0\textwidth]{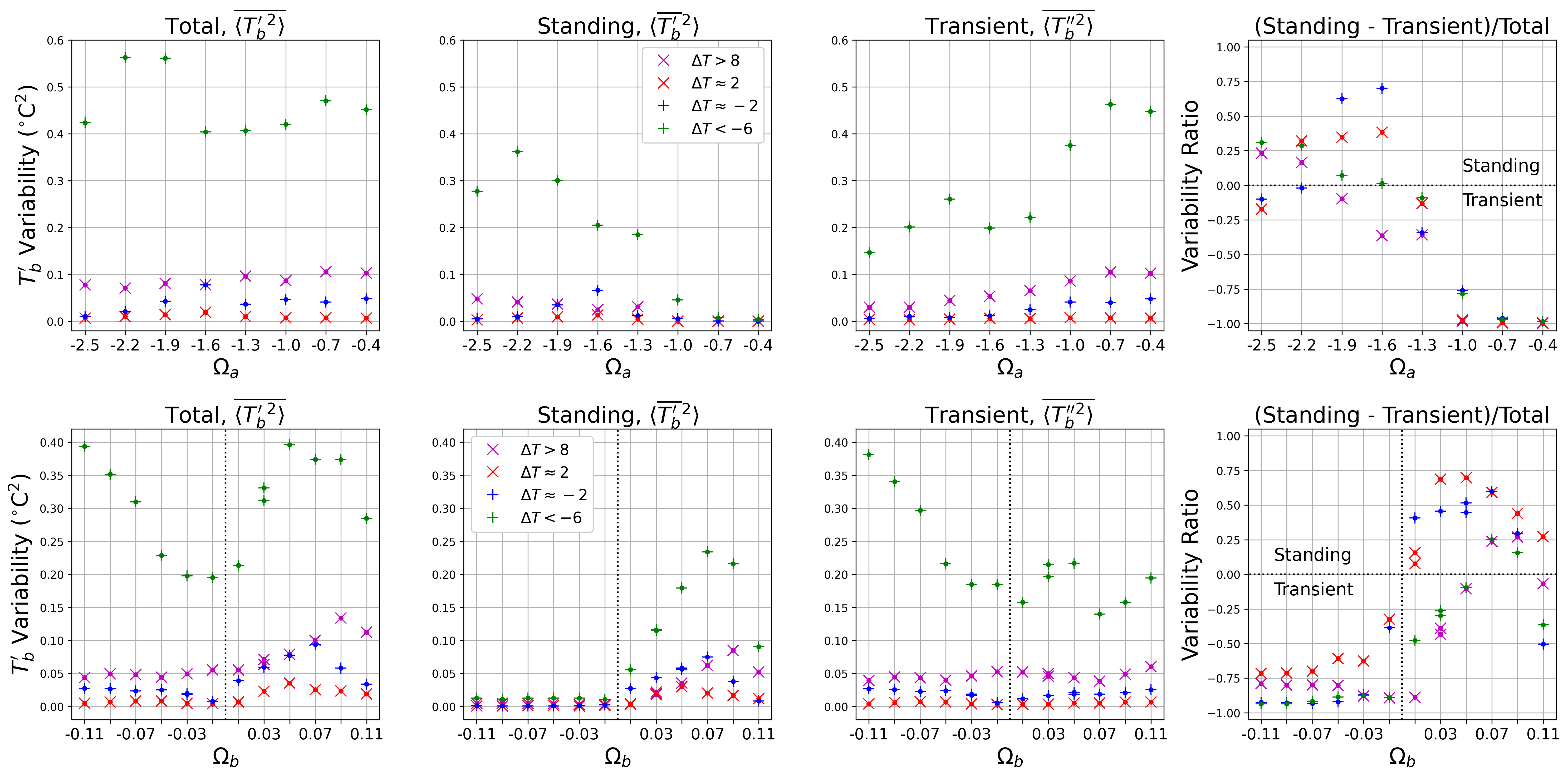}
\caption{Comparisons of the total, standing and transient variabilities (columns 1--3, respectively) for the set of experiments spanning annulus rotation rate $\Omega_{a}$ (top row) and bump rotation rate $\Omega_{b}$ (bottom row). The colours of the data points represent the different sidewall temperature differences $\Delta{T}$, with positive and negative $\Delta{T}$ cases shown as crosses and plusses, respectively. The 4th column shows the variability ratios of the experiments, given as the difference between the standing and transient variabilities normalised by the total variability; a variability ratio of -1 (+1) indicates the total variability is dominated by transient (standing) variability, and 0 indicates the total variability is equipartitioned between the standing and transient variability components.}
\label{fig:11_energetics}
\end{figure}

The spatial mean total thermal variability ($\overline{\langle{T^{\prime}_{b}}^{2}\rangle}$) in each experiment can be partitioned into a component which is standing or stationary relative to the bump (${\langle{\overline{T^{\prime}_{b}}}^{2}\rangle}$), and a component which is transient ($\overline{\langle{T^{\prime\prime}_{b}}^{2}\rangle}$), as expressed in Equation \ref{eqn:energy_terms}.
Figure \ref{fig:11_energetics} shows the total, standing, and transient thermal variabilities (columns 1--3, respectively) for experiments with varying annulus rotation rate $\Omega_{a}$ (top row) and varying bump rotation rate $\Omega_{b}$ (bottom row).
The experiments with larger sidewall temperature differences $\Delta{T}$ exhibit relatively more thermal variability than their smaller $\Delta{T}$ counterparts; this is to be expected as these experiments have a relatively wider range of temperatures in the working fluid.
Furthermore, the negative $\Delta{T}$ cases have substantially more variability than their corresponding positive $\Delta{T}$ cases; this is indicative of the positive $\Delta{T}$ cases being relatively more susceptible to baroclinic instability and thus mixing, the latter of which acts to reduce the range of temperatures in the working fluid, and thus the thermal variability.
In terms of the effect of annulus rotation rate $\Omega_{a}$, the total variability in each experiment appears relatively insensitive to $\Omega_{a}$ and without an overall trend; there are slight decreases in total variability as the annulus rotation rate goes from $\Omega_{a}=-0.4$ to $\Omega_{a}\approx-1.3$ and from $\Omega_{a}=-1.6$ to $\Omega_{a}=-2.5$, with a sharp increase in between these ranges.
The standing variability generally increases as the magnitude of $\Omega_{a}$ increases, with a marked increase at $\Omega_{a}\approx-1.3$\,rad/s.
By contrast, the transient variability decreases as the magnitude of $\Omega_{a}$ increases.

The fourth column of Figure \ref{fig:11_energetics} shows the ratio of the difference between the standing and transient thermal variabilities to the total thermal variability, that is,
\begin{equation}
\frac{({\langle{\overline{T^{\prime}_{b}}}^{2}\rangle}-\overline{\langle{T^{\prime\prime}_{b}}^{2}\rangle})}{\overline{\langle{T^{\prime}_{b}}^{2}\rangle}}.
\end{equation}
Recall that $\overline{\langle{T^{\prime}_{b}}^{2}\rangle} = \left({\langle{\overline{T^{\prime}_{b}}}^{2}\rangle} + \overline{\langle{T^{\prime\prime}_{b}}^{2}\rangle}\right)$, such that a variability ratio of -1 indicates the total thermal variability is entirely transient, a ratio of 1 means the total is entirely standing, and a ratio of 0 means the total is equipartitioned between standing and transient.
The variability ratio is near -1 for slow annulus rotation rates ($|\Omega_{a}|<1.0$\,rad/s); as the magnitude of $\Omega_{a}$ increases to 1.3\,rad/s the variability ratio tends toward 0, and becomes positive for most cases of $|\Omega_{a}|\geq1.6$\,rad/s.
That is, for these faster rotation cases, more than 50\% of the total thermal variability is contained in the standing variability fields.

A similar analysis is performed for the experiments different bump rotation rate $\Omega_{b}$ (bottom row).
These experiments exhibit the same overall sensitivity to the sidewall temperature differences $\Delta{T}$; larger $|\Delta{T}|$ cases exhibit more thermal variability, and negative $\Delta{T}$ cases have more thermal variability than their positive $\Delta{T}$ counterparts.
In general, the eastward flow cases ($\Omega_{b}>0$) exhibit more total thermal variability than their westward flow counterparts.
The eastward flow cases have standing variability that increases with the magnitude of $\Omega_{b}$ up to $\Omega_{b}\approx0.07$\,rad/s, beyond which it begins to decrease; by comparison, the westward flow cases have negligible standing thermal variability for all bump rotation rates.
Transient variability is comparatively less sensitive to bump rotation rate, except in the case of large negative $\Delta{T}$ wherein the transient variability increases with $|\Omega_{b}|$ for westward flow experiments.
The variability ratio clearly demonstrates that the thermal variabilities of the westward flow cases ($\Omega_{b}<0$) are dominated by their transient components, while eastward flow cases tend to have greater standing variability contributions.
For these eastward flow cases, there is a clear distinction between the large and small $|\Delta{T}|$ experiments; smaller $|\Delta{T}|$ cases have proportionally more standing variability than their larger $|\Delta{T}|$ counterparts, especially for the range $0.01<\Omega_{b}<0.07$.
The variability ratio tends to decrease as the magnitude of $\Omega_{b}$ increases beyond $\Omega_{b}=0.09$\,rad/s.

\begin{figure}
\centering
\includegraphics[width=1.0\textwidth]{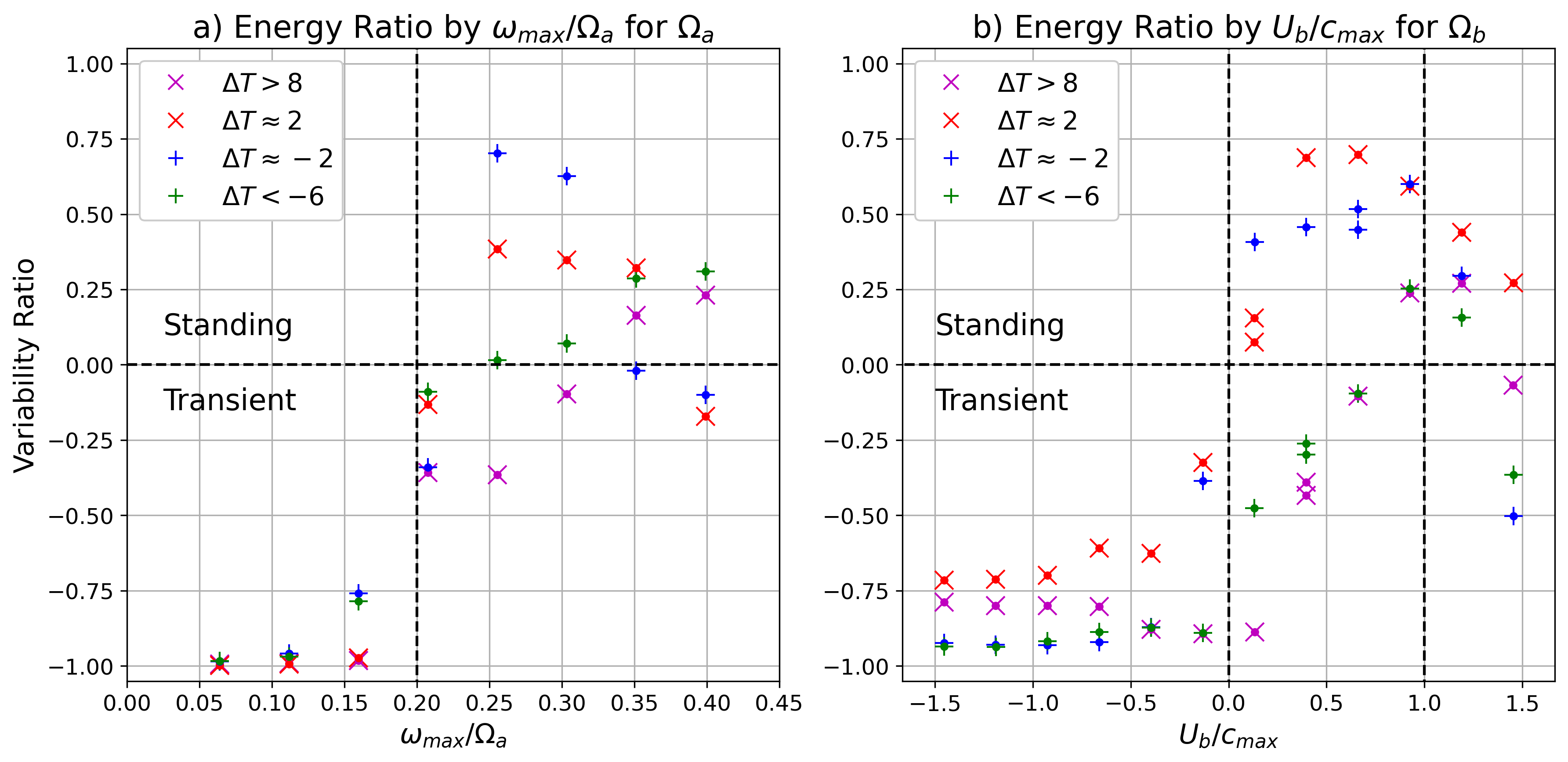}
\caption{The variability ratios plotted against (a) the ratio of the maximum barotropic Rossby wave frequency to the annulus rotation, $\omega^{BT}_{max}/\Omega_{a}$, and (b) the ratio of the maximum baroclinic Rossby wave phase speed to the background flow, $U_{b}/c_{max}$. Panel (a) only shows the set of experiments that vary the annulus rotation rate $\Omega_{a}$; panel (b) only shows the set of experiments that vary the bump rotation rate $\Omega_{b}$. Data point colours and symbols are as per Figure \ref{fig:11_energetics}.}
\label{fig:12_energetics_explained}
\end{figure}

The behaviours of the variability ratios in Figure \ref{fig:11_energetics} are indicative of transitions between distinct dynamical regimes relating to the annulus rotation rate and the bump rotation rate.
For the first, one hypothesis is that the background radial gradient in potential vorticity crosses a dynamical threshold such that the system is able transfer baroclinic energy from the background geostrophic turbulence into the barotropic mode and drive an inverse cascade to scales larger than the baroclinic Rossby deformation radius \cite[e.g.,][]{galperin_etal2006, galperin_etal2008}.
To test this, we compare the variability ratio with the ratio of the maximum Rossby wave frequency to the annulus rotation, $\omega_{max}/\Omega_{a}$.
Figure \ref{fig:12_energetics_explained}a shows the set of experiments with varying annulus rotation rate $\Omega_{a}$ and constant bump rotation rate $\Omega_{b}=0.05$\,rad/s.
The variability ratio in experiments with $\omega_{max}/\Omega_{a}<0.2$ is less than -0.75, indicating that they are dominated by transient variability.
For experiments with $\omega_{max}/\Omega_{a}>0.2$, the variability ratio increases and becomes positive for $\omega_{max}/\Omega_{a}>0.25$.

It is worth reflecting on the parallels between this ratio of the maximum Rossby wave frequency to the annulus rotation rate and the planetary number $\beta_{1}$ (Eqn. \ref{eqn:beta1}).
Specifically,
\begin{equation}
\frac{\omega_{max}}{\Omega_{a}}=\frac{\beta{L_{R}}}{2\Omega_{a}}\quad:\quad\beta_{1} = \frac{\beta{L}}{2\Omega_{a}},
\end{equation}
such that the $\omega_{max}/\Omega_{a}$ ratio is simply the planetary number with the lengthscale $L$ substituted for the barotropic Rossby deformation radius.
The dynamical implication of this here is that experiments which have a relative change in their effective Coriolis parameter of more than $20$\% over the barotropic Rossby deformation radius will be able to support variability that is non-transient.
Equivalently, there exists a threshold of the maximum Rossby wave frequency relative to the background rotation frequency for which variability is able to become non-transient; that is, if the maximum Rossby wave frequency supported by the system is too low relative to the background rotation frequency, variability can only be transient.
Recall that the Earth-like mid-latitude atmospheric regime spans $0.2<\beta_{1}<1.0$ (Fig. \ref{fig:02_parameterspace}c).

In terms of the bump speed, there is a clear distinction between the eastward versus westward flow cases; westward flow cases ($\Omega_{b}<0$) are dominated by transient variability, while eastward flow cases have predominantly standing variability.
The variability ratios of the eastward flow cases do suggest there are additional processes at play, in particular the decrease in the variability ratio as the bump becomes faster.
In Figure \ref{fig:12_energetics_explained}b we compare the variability ratio with the ratio of the bump velocity to the maximum Rossby wave phase speed, $U_{b}/c_{max}$.
This ratio can be thought of as a Froude or Mach number; the ratio of a flow speed to a wave speed.
While the phase speed is positive definite, the bump velocity is not, such that the ratio spans positive and negative values, with positive values indicative of eastward flow cases.

There is a sudden increase in the variability ratio as $U_{b}/c_{max}$ increases from negative to positive.
The increase continues until $U_{b}/c_{max}$ passes through unity, beyond which the variability ratio decreases.
The regime between $0<U_{b}/c_{max}<1$ is that in which the flow relative the bump is both eastward and slower than the maximum Rossby wave phase speed; that is, the system is sub-critical in terms of Rossby waves.
Increasing the flow speed to the super-critical regime of $U_{b}/c_{max}>1$ transitions the system back to a predominantly transient variability state.
Note that the cases with larger $|\Delta{T}|$ tend to have relatively smaller variability ratio than their smaller $|\Delta{T}|$ counterparts.
Again, considering the equation for the maximum Rossby wave speed (Eqn. \ref{eqn:speedmax}), it is worth reflecting on the dynamical similarities between this ratio and the form of the expression for the planetary-relative vorticity number $\beta_{2}$ (Eqn. \ref{eqn:beta2}):
\begin{equation}
\frac{U_{b}}{c_{max}}=\frac{U_{b}}{\beta{L_{R}^{2}}}\quad:\quad\beta_{2} = \frac{\beta{L^{2}}}{U}.
\end{equation}
That is, the speed ratio we are using here can be though of as being representative of the inverse of the planetary-relative vorticity number $\beta_{2}$ where the lengthscale $L$ is substituted for the barotropic Rossby deformation radius, and the velocity scale is the (eastward) background flow speed.
Systems that have westward flow or super-critical eastward flow will tend to have predominantly transient variability; systems with sub-critical eastward flow can support standing variability.

\section{Conclusions}

The experiments presented here span a wide range of dynamical parameter space including that which is relevant for investigating mid-latitude atmospheric dynamics, such as standing Rossby waves.
Overall, the system behaves in an intuitive and predictable manner; increasing the annulus rotation rate strengthens the background gradient of potential vorticity sufficiently to support Rossby waves, which are able to become standing only in cases with eastward flow.
The wavelength of the standing Rossby waves depends on the speed of the eastward flow and inversely on the strength of the potential vorticity gradient.
The sidewall temperature difference governs the baroclinicity of the system; configuring the experiment such that the slopes of the isopycnals and isobars are opposed enhances the tendency for baroclinic instability, and a therefore more mixed state.
In addition to successfully identifying the experimental configurations for standing Rossby waves, the exercise here has established and explained the bounds for these parameters; for example, forcing the system with eastward flow that is faster than the Rossby wave phase speed leads to a deterioration of the wave structures and a shift back to predominantly transient variability.
Future efforts will build these findings to explore the characteristic response of standing Rossby wave behaviour to projected changes in Earth's climate, in particular, the influence of polar amplification.

\section*{Acknowledgments}
We wish to thank Angus Rummery, Peter Lanc, and Tony Beasley for the construction of the apparatus, technical support, and laboratory assistance.
We acknowledge very helpful discussions with Marty Singh, Michael Barnes, Julie Arblaster, Michael Reeder, Martin Jucker, and Navid Constantinou, and are indebted to the academic brilliance of Ross Griffiths.
TGS was supported by an {\it{Australian Research Council Centre of Excellence for Climate Extremes}} Undergraduate Research Scholarship.

\clearpage

\bibliography{stewart_etal_GAFD.bib}
\bibliographystyle{gGAF}

\end{document}